\documentclass[twocolumn]{aa}  

\usepackage{graphicx}
\usepackage{txfonts}
\usepackage{natbib}
\bibpunct{(}{)}{;}{a}{}{,}
\begin{document} 

\def \CO12 {$^{12}$CO\, }
\def \13CO {$^{13}$CO\, }

  \title{High-spectral resolution M-band observations of CO Rot-Vib absorption lines towards the Galactic center}

  \subtitle{}

  \author{J. Moultaka
         \inst{1},
          A. Eckart
          \inst{2,3},
          K. Tikare
          \inst{4}
          \and
         A. Bajat  
           \inst{5}        
          }

   \institute{
1) IRAP, Universit\'e de Toulouse, CNRS, CNES, UPS, 
 14, avenue Edouard Belin, F-31400 Toulouse, France,
      \email{jihane.moultaka@irap.omp.eu}\\
      2) I Physikalisches Institut, Universit\"at zu K\"oln, Z\"ulpicher Str. 77, D-50937 K\"oln, Germany,
    \email{eckart@ph1.uni-koeln.de}\\
      3) Max-Planck-Institut f\"ur Radioastronomie, Auf dem H\"ugel 69, D-53121 Bonn, Germany\\
4) Lulea University of Technology, Kiruna, Sweden\\
5) Institute of Physics of the Czech Academy of Sciences, Prague
            }

   \date{Received ; accepted }

 
  \abstract
   {In the near- to mid-infrared wavelength domain, bright continuum sources in the central parsec of the Galactic center (GC) are subject to foreground absorption. These sources therefore represent ideal probes of the intervening material that is responsible for the absorption along the line of sight.}
   { Our aim is to shed light on the location and physics of the absorbing clouds. We try to find out which of the gaseous absorbing materials is intimately associated with the GC and which one is associated with clouds at a much larger distance.}
   {We used the capabilities of CRIRES spectrograph located at ESO Very Large Telescope in Chile to obtain absorption spectra of individual lines at a high spectral resolution of R=65000, that is, 5~km/s. We observed the \CO12 R(0), P(1), P(2), P(3), P(4), P(5), P(6), P(7) and P(9) transition lines, applied standard data reduction, and compared the results with literature data.}
   {We present the results of CRIRES observations of 13 infrared sources located in the central parsec of the Galaxy. The data provide direct  evidence for a complex structure of the interstellar medium along the line of sight and in the close environment of the central sources. In particular we find four cold foreground clouds at radial velocities v$_{LSR}$ of the order of -145,-85, -60, and -40$\pm$15 km/s that show absorption in the lower transition lines from R(0) to P(2) and in all the observed spectra. We also find in all sources an absorption in velocity range of 50-60 km/s, possibly associated with the so-called 50~km/s cloud and suggesting an extension of this cloud in front of the GC. Finally, we detect individual absorption lines that are probably associated with material much closer to the center and with the sources themselves, suggesting the presence of cold gas in the local region.
}
   {}


\titlerunning{CRIRES observations of the GC}
\authorrunning{Moultaka et al.}

 \maketitle
%
%
\section{Introduction}
\begin{figure*}
\includegraphics[width=40pc,angle=0]{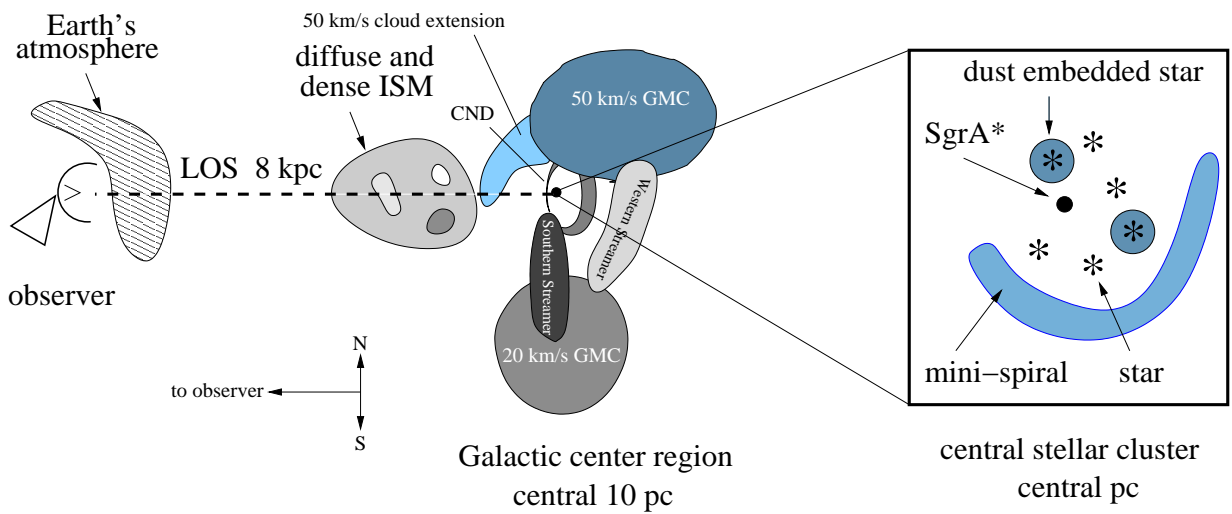} 
\caption{\label{LOS} 
Schematic view explaining the origin of absorption lines towards the GC. Part of the absorption is taking place in the Earth's atmosphere (left). This is usually calibrated using reference stars. However, residuals may still be present or genuine astrophysical absorption lines may be difficult to 
access if they fall onto an atmospheric line. Part of the absorption may take place in the cold, dense, and diffuse ISM in the LOS towards the GC (middle, \citet{whittet97}). Most of this absorption probably takes place within the final few hundred parsecs. \citet{goto14} point out that the absorptions cover a wide range of velocities and LOS locations within the  central molecular zone close to the central stellar cluster. Within the central 10~pc there are a few molecular cloud complexes that come quite close to the LOS (right). Following \citet{lee08} some of them are indicated here. The 50~km/s cloud has been indicated in light blue. See comments on the 50~km/s extension in the caption of Fig.~\ref{leefig}. Finally, individual absorptions in the local environment of the central cluster may take place (right inset). This may happen against the minispiral or against the bowshocks, shells, and circumstellar envelopes of some or the MIR excess sources in the central cluster. Some representative structures have been indicated in the inset.
}

\end{figure*}

The central parsec of our Galaxy, often referred to as the Galactic center (GC), is a very intriguing region that has been studied for decades. It harbors a supermassive black hole, Sgr~A$^{\star}$ (\citet{schoedel03}, \citet{ghez03}), surrounded by a dense central cluster where late-type red giants, young massive stars, and bowshock shaped infrared (IR) sources are identified (\citet{ott99}, \citet{paumard04}, \citet{perger08}, \citet{tanner03}, \citet{eckart04}). 
The interstellar medium (ISM) in the region of the GC consists of several components: In the very central parsec, it shows a spiral feature called "the minispiral" (or Sgr~A West) composed of dust and ionized gas where thin dust filaments are also observed, probably resulting from interactions with stellar winds (e.g., \citet{muzic07}, \citet{paumard01}). Furthermore, the GC is surrounded by the circumnuclear disk (CND), a molecular ring of dense gas and dust (\citet{guesten87}, \citet{garciamarin11}, \citet{mossoux18}). These structures are responsible in part for the very high extinction observed towards the GC (about $\sim 35$~mag in the visible; e.g., \citet{schoedel07}, \citet{scoville03}). The main obscuration however is due to the diffuse ISM and dense molecular clouds present along the line of sight (LOS). Only one third of it would be due to foreground dense molecular clouds (\citet{whittet97}).
Also, in our previous works (\citet{moultaka04}, \citet{moultaka05}, \citet{moultaka15a}) we showed that local absorptions probably occur in the direct environment of the GC IR sources. As a matter of fact, the spectra of luminous IR sources in the L-band (from 2.7 to 4.2$\mu$m) revealed absorption features due to water ices and hydrocarbons that we claimed to be present in the local medium of the central parsec. These absorptions were attributed earlier by other authors to material in the foreground molecular clouds (e.g., \citet{chiar00}, \citet{chiar02}).\\
In Fig.~\ref{LOS}, we show the complex observational situation with respect to absorption lines that occur along the LOS towards the GC. The absorptions can occur at several locations, some of which may be associated with the immediate environment of the stellar cluster and the ISM structure close to or within it.

More recently, we used the ISAAC ESO/VLT spectrograph capabilities in the low-resolution (R=800) mode of the M-band wavelength range (from 4.4 to 5.1$\mu$m ) to study the spectra of 15 bright GC sources (\citet{moultaka09}) and to map a significant region of the central parsec in the M-band (\citet{moultaka15b}). 
This resulted in the high-quality M-band spectra ($<$0.8'' resolution and high S/N) and in a larger number of GC sources than previously published works (\citet{lacy84}, \citet{tegler93}, \citet{geballe86}, \citet{mcfadzean89}, \citet{geballe89}, \citet{mon01}). Moreover, the fundamental P and R branches of the ro-vibrational transition lines of the \CO12  and \13CO  gas-phases (around 4.666$\mu$m and 4.77$\mu$m, respectively) were resolved in the GC. We found that an appreciable amount of the CO in the foreground ISM may also originate from solid phase, stuck to the surface of dust grains, and that both gas and dust of the diffuse medium are responsible for a major part of the extinction.
We performed a first-order correction of individual GC source spectra for the foreground absorption due to CO-ice and gas. The resulting corrected spectra for the CO-ice absorption at 4.675$\mu$m revealed residual absorption in a number of sources. The residual absorption may be due to variations in the foreground LOS extinction. In general, all previous works concluded that it is unlikely that the molecular clouds are close to the GC because the CO ice is unlikely to survive temperatures higher than 20~K unless it is located in an environment of water ice and other molecules (e.g., \citet{mcfadzean89}). We claimed that part of the residual CO-ice may be due to GC material (i.e., minispiral) or to material within the circumstellar shells or bow shocks of the sources. We showed that given the structure of the GC and  the presence of water ices (\citet{moultaka04}, \citet{moultaka05} and \citet{moultaka15a}), bowschock sources, and dust filaments, CO-ice may survive in this environment.
Moreover, we observed a residual CO-gas absorption in the corrected spectra. Assuming that the residual gas material is located in the minispiral or in the circumstellar shells of the dust-embedded bow-shock sources, we obtained gas masses of the circumstellar shells of the order of 10$^{-3}$M$_{\odot}$ in agreement with an independent estimate of circumstellar shell masses using published CO-(7-6) and FIR[OI] data.
Consequently, we deduced that a substantial part of the 4.666$\mu$m gaseous CO absorption could be due to intrinsic circumstellar material, although \citet{geballe89} attributed the CO R(2) and R(5) lines towards IRS~1, 2, 3, 5, 6, 7, and 8 GC sources observed at velocities between 0 and 75~km/s to the gas in the spiral arms along the LOS to the GC and to the 20~km/s and 50~km/s known foreground clouds close to the GC. \\
If this is the case and if our conclusion/assumption was correct, the CO absorptions would occur at the radial velocities of the sources, many of which are larger than the 20~km/s and 50~km/s cloud velocities (see Table~\ref{tabvel}).\\

In this paper, we present the results of our spectroscopic observations\footnote{Based on observations collected at the European Southern Observatory under ESO programme 093.C-0086(A).} of individual IR sources of the central parsec at a higher spectral resolution (R=65000 i.e., 5 km/s) than what we have previously accomplished, using the capabilities of CRIRES spectrograph located at ESO Very Large Telescope. Our goal is to identify intrinsic gaseous CO absorption by checking if the CO lines occur at the radial velocities of the sources and consequently if the CO is located in the circumstellar medium of these sources. In the following section, we describe the observation setups and the data-reduction steps. In Sect.~3, we show and discuss our results in terms of local and foreground absorptions and finally we conclude in the last section.

\section{Observations and data reduction}\label{datareduction}

\begin{figure}
\includegraphics[width=15pc]{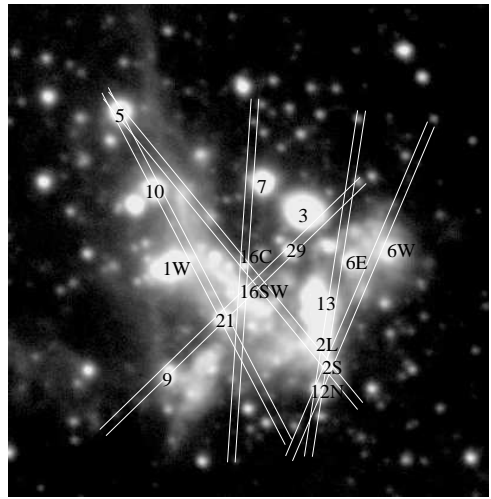}
\caption{\label{Slits} Positions of the 0.4$\arcsec$ slits used during the observing run. They cover the 13 observed sources listed in Table~\ref{tabvel}. The IRS numbers of the main bright sources of the central cluster are also indicated.}
\end{figure}

The observations were obtained during period 93\footnote{Observing program run 093.C-0086(A)} on three half nights (18, 19, and 20 May, 2014) with CRIRES high-resolution spectrograph located at UT3 of the VLT/ESO observatory. We used a 0.4$\arcsec$ slit in the high-resolution mode of CRIRES which results in a spectral resolution of about 65,000.\\
For our scientific goals, we observed at two reference wavelengths, 47233 and 47363\AA, according to the definition of the CRIRES manual (i.e., the reference wavelength is the one at pixel 512 of the third detector). 
At each of the reference wavelengths, the observations were obtained in four distinct ranges that do not overlap, corresponding to the four detectors of the instrument. In Table~\ref{tablambda} we show the obtained ranges for each reference wavelength. On average, the exposure time was about 1.5 hours per slit position and instrument setup. In total, six slit positions (see Fig.\ref{Slits}) were used either with the 47233\AA\ reference wavelength setup or the 47363\AA\ one. In each of the slit positions a number of IR GC sources are observed. This resulted in data for 13 IR sources of the GC (in one or both reference wavelength settings): IRS 1W, 2L, 2S, 3, 5, 6W, 10W, 12N, 13, 16C, 16SW, 21, and 29.\\
The observations were performed using the nodding and jittering techniques. Nodding consists of moving the telescope from one position to a second one along the direction of the slit and then moving it again to the first position. Jittering is a small offset that is applied randomly in addition to the nodding to avoid systematic errors and to correct for bad pixel detectors. \\
Thanks to the nodding technique, the dark current correction and sky emission removal are automatically obtained by subtracting the two offset images. Using the ESO CRIRES pipeline, the data were dark subtracted, corrected for nonlinearity, flatfielded and each couple of nodded images were combined. These steps are obtained when executing the crires-spec-jitter recipe of the ESO pipeline. At that step, the pipeline also extracts the brightest spectrum in the image and applies the wavelength calibration. In the present observations, several spectra are present in a single image since we positioned the slit along a series of IR sources of the GC (see Fig. \ref{Slits}). Moreover, the traces of the spectra in the detector images are curved. Thus, after making different unsuccessful tests to extract the spectra from one detector image and to straighten them using the pipeline recipe, we decided to use our own scripts, written with the IRAF facility\footnote{IRAF is written and supported by the National Optical Astronomy Observatories (NOAO) in Tucson, Arizona. NOAO is operated by the Association of Universities for Research in Astronomy (AURA), Inc. under cooperative agreement with the National Science Foundation}. Our scripts allow for the extraction process to be performed after straightening the traces of the spectra in the detector images. To this end and prior to this step, we extracted the four detector images from the fit output files of the crires-spec-jitter recipe and saved each of them in a one-extension FITS file. The output FITS files of the CRIRES pipeline recipe have four extensions where the four detector images are saved, respectively. The wavelength calibration function was obtained using the sky emission lines present in the observations as well as observed lines from an N$_2$O gas-cell that was positioned in front of a halogen lamp as part of the ESO standard calibration plan. To do the wavelength calibration, we used the polynomial relation (between pixel positions and wavelengths) for each of the detector images obtained with the CRIRES ESO pipeline and applied it to the extracted spectra of each of the detector images.\\
The last step of the data-reduction process was the removal of telluric lines and flux calibration. This step was very difficult to perform although we observed a telluric standard soon after and before each of the science exposures using the same instrument setup and at airmasses very close to the ones obtained during the science observations. To this end, we performed several tests with the Molecfit software tool (\citet{smette15} and \citet{kausch87}) dedicated to the correction of telluric absorption lines using synthetic atmospheric spectra. Most of the results were not convincing, and therefore for most of our observed spectra we performed a telluric correction by hand using our IRAF scripts and the observed telluric standard spectra. The observed telluric stars are the early-type A0V stars HR~6070, HIP~117089, and B9IV star HR~6494. The final spectra, shown in black in Figs.~\ref{IRS1W} to \ref{IRS16C} are not ideally corrected, as can be seen in the figures, but they are the best results that we were able to obtain. Therefore, in each of the figures presenting a science reduced spectrum, we also show the corresponding telluric star spectrum to assess our discussion and results. In general, all the wavelength ranges lying on a region of a telluric line are left out of the discussion.
\begin{table*}
\begin{center}
\begin{tabular}{ccc}
\hline\hline
Reference wavelength & 47233 & 47363 \\ 
       &  (\AA)    &  (\AA) \\
\hline
wavelength range in detector 1  & 46519-46761  & 46656-46896 \\
wavelength range in detector 2  & 46828-47061  & 46962-47192 \\
wavelength range in detector 3  & 47122-47344  & 47253-47473 \\
wavelength range in detector 4  & 47401-47613  & 47529-47739 \\
\hline
\end{tabular}
\label{tablambda}
\caption{The wavelength ranges (in Angstr{\"o}m) of the spectra obtained at each of the CRIRES reference wavelengths. }
\end{center}
\end{table*}

\begin{table*}
\begin{center}
\begin{tabular}{lcccc}
\hline\hline
Source  & type & Radial velocity & reference                       & $v_{LSR}$ of a local absorption\\
        &      &  (km/s)    &                                      & (km/s)  \\
\hline
IRS 1W  & red  & $35\pm20$  & Paumard et al. (2006)                &$35\pm15$ in R(0)\\
        &      &            &                                      &$50\pm15$ in P(1)\\
IRS 2S  & cool & $107\pm20$ & Genzel et al. (2000)                 &-$110\pm15$ in R(0) P(1) and P(2)\\
IRS 2L  & red  &      -      &                                      &$60\pm15$ in R(0), P(1), P(2) and P(3)\\
        &      &            &                                       &{\bf-105}$\pm${\bf15} in R(0), P(1) and P(3)\\
IRS 3   & red  &      -      &                                      &{\bf60}$\pm${\bf15} in R(0),P(1),P(2),P(3),P(4),P(6) and P(7)\\
IRS 5   & cool & $110\pm60$ & Tanner et al. (2005) (flow)            &$70\pm15$  in P(1), P(2), and P(3)\\
IRS 6W  & hot  & $63\pm3$   & Zhu et al. (2008), Herbst (1993)     &$-20\pm15$ in P(1), P(2), P(4) and P(5)\\
        &      &            &                                      &$65\pm15$ in P(1), P(2), P(4) and P(5) \\
IRS 10W & cool & $80\pm60$  &  Tanner et al. (2005) (source)       &$50\pm15$ in P(1) and P(2)\\
        &      &            &                                      &$-105\pm15$ in R(0), P(1), P(2) and P(3) \\
        & cool & $195\pm150$&  Tanner et al. (2005) (flow)         &$110\pm15$ in P(3)\\                            
        &      &            &                                      &$210\pm15$ in P(3)\\ 
IRS 12N & cool & $-69\pm2$  &  Zhu et al. (2008)                   &$-85\pm15$ in R(0), P(1),P(3),P(4),P(6),P(7) and P(9)\\
        &      & $-61\pm1$  &  Figer et al. (2003)                  &\\
IRS 13E & hot  & $45\pm60$  &  Genzel et al. (2000)                &$60\pm15$ in R(0), P(1),P(2) and P(3)\\
IRS 16C & hot  & $125\pm30$ &  Paumard et al. (2006)               &-\\
IRS 16SW& hot  & $453\pm4$  &  Martins et al. (2006)               &-\\
IRS 21  & hot  & $-90\pm20$ &  Genzel et al. (2000)                &$-60\pm15$ in P(3)\\
        &      &            &                                      &$55\pm15$ in R(0), P(1), P(2) and P(3)\\
IRS 29S & hot  & $-93\pm20$ &  Genzel et al. (2000)                &$-60\pm15$ in P(3)\\
        & hot  & $-164\pm2$ &  Zhu et al. (2008)                   &\\
\hline
\end{tabular}
\label{tabvel}
\caption{Radial velocities of the individual observed sources (in column 3) as found in the literature (references are given in column 4).  
The second column shows the spectral type of the sources listed in column 1.
Most of the radial velocities clearly lie beyond the 20-50~km/s cloud velocities.
For sources IRS5 and IRS10W, \citet{tanner05} give separate values for the source and the
corresponding minispiral flow (see e.g., velocity field given by \citet{zhao10}). For IRS10W, two `flow' values are given ($175\pm130$~km/s and $215\pm160$~km/s) of
which we list the average in the table.
Sources IRS1W, IRS2L, and IRS3 are enshrouded in dust and are therefore labeled as `red'.
For IRS2L and IRS3 we found no quoted velocity.
For IRS6W, we classify this source as "hot" using the information by \citet{herbst93} that the Br$\gamma$ line emission is present in its spectrum. The radial velocity
given by \citet{herbst93} has not been tabulated in their paper but can be read off their Fig. 5 as
-166$\pm$30~km/s and may be associated with the gas flow in the minispiral bar. 
In column 5, we present the assumed local gas velocities measured in our spectra in the CO transition lines indicated. The values in bold are the first determinations of the radial velocities of IRS~2L and IRS~3 that we derive from the present work (see Sect.\ref{localabs}).
}

\end{center}
\end{table*}

\section{Results}
As explained in the previous section, most of our spectra show residual telluric lines that were very difficult to remove during the data-reduction process. In the following, all the results we present concern \CO12 lines in spectra for which we are confident that the features are real, that is, not telluric. In Figs.~\ref{IRS1W} to \ref{IRS16C} one can see examples of observed reduced spectra with the corresponding observed telluric standard stars shown in red. Thereby, we do not discuss the lines in the science spectra that fall at the location or close to the location of the telluric lines. One has to be aware that if we do not mention the presence of a line at a given velocity in a spectrum, this does not necessarily mean that there is no line but it could be that at that velocity, a residual telluric line is present, meaning that we cannot provide any conclusion on its presence in the observed science spectrum.\\
Moreover, in the following discussion, it is important to note that the absorption lines at low transitions, like R(0), P(1), and P(2) for example, are indicative of a low temperature in the medium where they are produced. On the contrary, high-absorption transition lines indicate higher temperatures of the medium. Therefore,  the absorption lines in P(3) to P(7) that we observe could be associated with the very local medium surrounding the individual sources.\\
Finally, we also note that R-branch, low J lines of \13CO are also present in the wavelength intervals observed here but these absorptions do not contaminate any of the velocity profiles of the \CO12 R- or P-branch lines that we present.

\subsection{Absorption along the line of sight\label{sectionLOS}}
\begin{figure}

\includegraphics[width=20pc,angle=0]{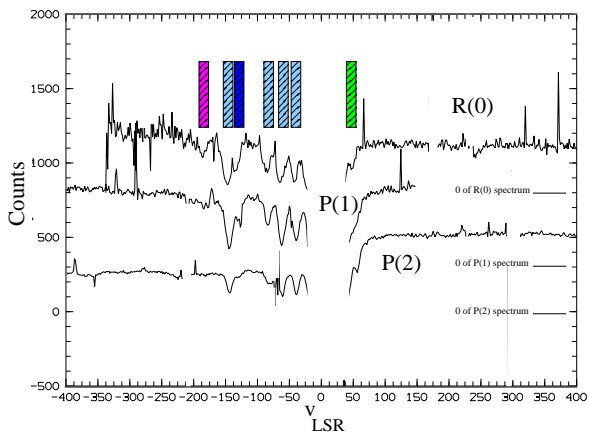}  
\caption{\label{sumR0_P1_P2} Spectra of the \CO12  R(0), P(1), and P(2) transition lines resulting from the addition of all the observed IR source spectra. The light blue rectangles show the positions of the lines at $v_{LSR} \sim -40, -60, -85,$ and $-145$km/s present in all the spectra and that we attribute to foreground clouds. The magenta, blue, and green rectangles show the positions of the lines at velocities $v_{LSR}\sim$-185, -130, and +50~km/s, respectively. They are present in two of the transition line spectra (R(0) and P(1) in the case of the lines at $v_{LSR}\sim$-130 and -185~km/s, and R(0) and P(2) in the case of the line at $v_{LSR}\sim$50~km/s).}  

\end{figure}

\begin{figure}

\includegraphics[width=20pc,angle=0]{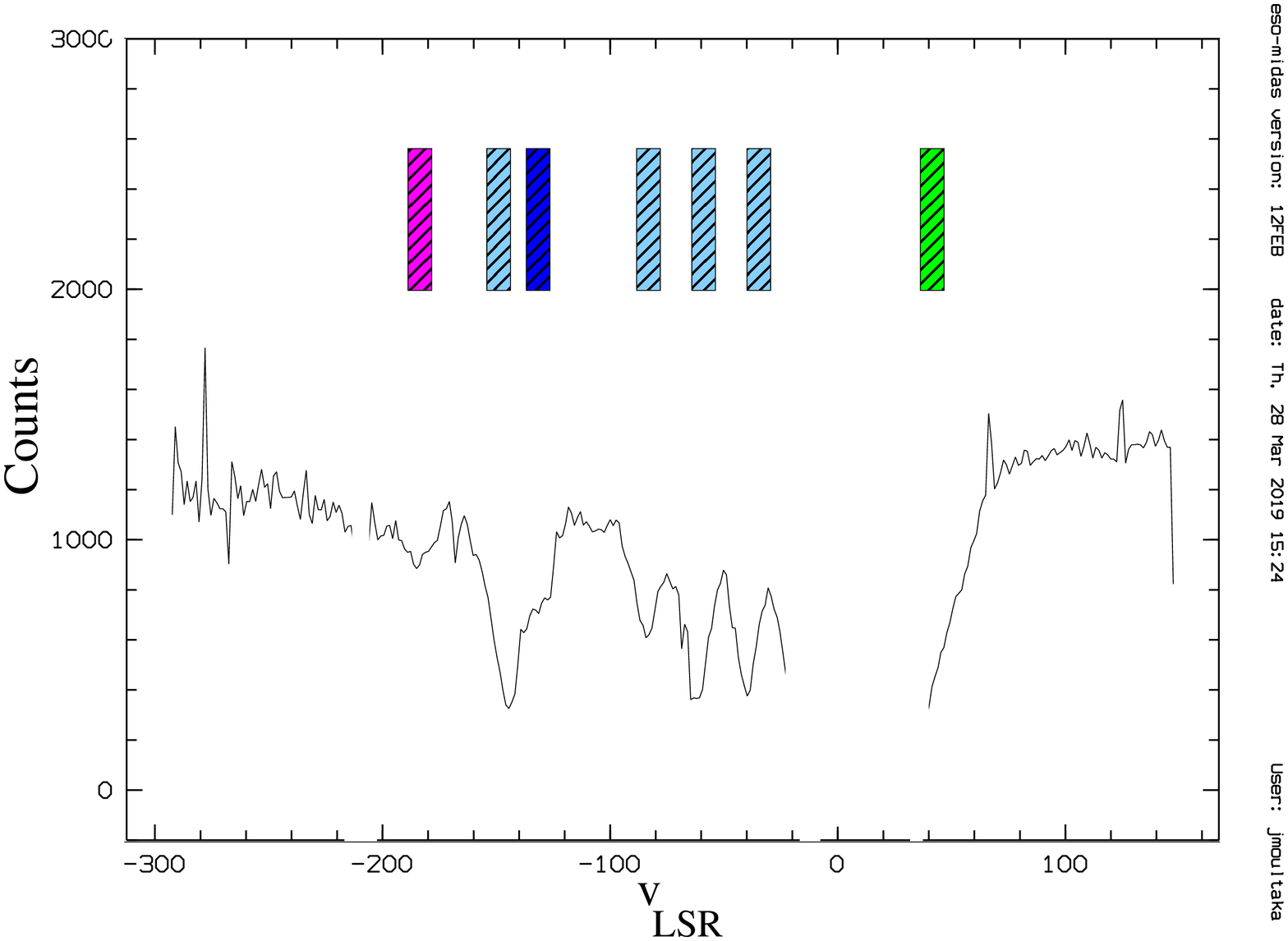}  
\caption{\label{sumR0P1P2} Resulting spectrum of all the summed observed GC IR source spectra in the \CO12  R(0), P(1), and P(2) line transitions (i.e., spectrum resulting from the three spectra shown in Fig.~\ref{sumR0_P1_P2}). The colored rectangles have the same indications as in Fig.~\ref{sumR0_P1_P2}. } 

\end{figure}

The individual spectra of the observed IR sources have different signal to noise ratios (S/Ns) depending on the integration time and brightness of the sources. We identified a set of \CO12  lines, especially in R(0), P(1), and P(2) transitions, located at LSR velocities ($v_{LSR}$) of $\sim$-40, -60, -85, and -145 km/s and another set of lines at -130 and -185 km/s. They are present in all or most of the spectra of the infrared sources (IRS) and are not (or are barely) contaminated by telluric lines (see Figs.~\ref{IRS1W} to \ref{IRS16C}). In order to highlight this result, we summed up all the IRS spectra for each of the \CO12 R(0), P(1), and P(2) line transitions and obtained the summed spectra shown in Fig.~\ref{sumR0_P1_P2}. We also show, in Fig.~\ref{sumR0P1P2}, the resulting spectrum obtained by adding the three previous spectra of all sources. In both figures, we distinguish the four negative velocities at $v_{LSR} \sim -40, -60, -85,$ and $-145$ km/s. These lines are most probably the signature of clouds located along the LOS. In the spectrum of the summed R(0) transition and that of the P(1) transition, we also distinguish two lines at $v_{LSR}\sim$-130 and -185 km/s that could be attributed to two other foreground clouds. Moreover, in the R(0) and P(2) transition summed spectra we find a faint absorption at around 50 km/s, 
possibly arising from the 50 km/s cloud at the GC; see, for example, \citet{wright01} and  \citet{uehara17}. \\ 
Finally, by comparing the velocities at the line peaks in the summed spectra between themselves and between the individual spectra, we estimate the error on our derived velocities to be of the order of 10 to 15 km/s.\\
In the following we describe the detections of the different lines that we assume to be absorbed in clouds located along the LOS but at different distances to the GC. 
The association of molecular absorption with individual sources of similar radial velocity may often be explained as absorption in a bow-shock of swept up local material (e.g., IRS1W or IRS5). Absorption feature originating in large-scale
extended shells or outflows are less likely to be expected in the harsh dynamical environment of the central stellar cluster - as these structures will be rapidly tidally destroyed. A source of some of the high-negative-velocity absorption components (between -100 and -200 km/s) could be due to CO in the expanding molecular ring or shell reported by \citet{kaifu72}, \citet{Scoville72} and \citet{Sofue95}. 
Others, for example \citet{Oka05}, attribute the high-negative-velocity absorption 
they observe in H$_3^+$ and the CO overtone band toward GCS3-2 to the ring.
A summary of the absorption line detections is given in Table~3 
 and all spectra are shown in Figs.~\ref{IRS1W} to \ref{IRS16C}.

\subsection{The case of IRS1W and IRS3}
In \citet{goto14}, the authors find absorptions of the \13CO v=1-0 P(1) and the $H_3^{+} R(1,1)^{l}$ transitions in the spectra of IRS~1W and IRS~3, at $v_{LSR}$ of -53, -32 km/s, and 0 km/s. They interpret the two absorbed features at -53 and -32 km/s as being produced in cold and dense gas of two lateral arms at 3 and 4 kpc moving at these velocities. These velocities were also observed towards the GC by \citet{mon01} and in radio by \citet{sutton90} and \citet{serabyn86}. Their finding is in agreement with the foreground velocities at -60$\pm$15 and -40$\pm$15 km/s that we observe in our summed spectra. We note that the systematic shift of 10 km/s between our measurements and those obtained in \citet{goto14} and the above references may be due to the wavelength calibration that was not straightforward during the data-reduction process as explained in Sect. \ref{datareduction}. In any case, this shift is inside the error bars that we obtain on the radial velocities (see Sect.\ref{sectionLOS}).\\
In the present individual spectra of IRS~1W and IRS~3, we find three absorptions of the \CO12 R(0) and P(1) lines at velocities of the order of $v_{LSR} =$ -60, -80, and -145 km/s even though the spectra are not of good quality (see Figs.~\ref{IRS1W} and \ref{IRS3}). An absorption at $\sim-40$~km/s is also visible in the \CO12 R(0) and the P(1) transition spectra of both IRS~1W and IRS~3 (see the figures), but there is a telluric line at that position. We note that most of these lines are saturated in these spectra. The absorption at 0 km/s found by \citet{goto14} was associated by the authors with gas located in foreground spiral arms. In our spectra, it can be marginally detected at low S/N in the \CO12 P(6) and P(7) lines of IRS~3 (see Fig.~\ref{IRS3}), but it falls together with a telluric line.\\

\begin{table*}
\begin{center}

\begin{tabular}{lcccccc}
\hline\hline
Source     & type                         & Stellar radial & Lines at  $v_{LSR}=$                                  & Lines at $v_{LSR}=$            & Lines at  $v_{LSR}=$         & Lines at  $v_{LSR}=$ \\
           &                              & velocity (km/s) &     -145,-85,-60, -40 km/s                             &    -185 km/s                          &    -130 km/s                        &   50 km/s\\
           &                              &                 &     observed in summed                    & observed in summed    &  observed in summed  &   observed in summed \\
           &                              &                 &    R(0), P(1) and P(2)             & R(0) and P(1)   &   R(0) and P(1) &    R(0) and P(2) \\
           &                              &                 &     spectra            &  spectra   &   spectra &     spectra \\
\hline 
Northern arm   & & & & & & \\
 sources  & & & & & & \\
\hline
IRS 1W      & red                          & $35\pm20$      & yes    & no    & yes   & yes \\
IRS 5       & cool                         & $110\pm60$     & yes    & yes   & yes   & yes\\
IRS 10W     & cool                         & $80\pm60$      & yes    & yes   & yes   & yes \\
            & cool                         & $195\pm150$    &        &       &       &    \\
IRS 21      & hot                          & $-90\pm20$     & yes    & yes   & yes   & yes\\
\hline
The Helium  &&&&&& \\
stars &&&&&& \\
\hline
IRS 16C       & hot                        & $125\pm30$     & yes    & no    & no    & yes\\
IRS 16SW      & hot                        & $453\pm4 $     & yes    & no    & no    & yes \\
\hline
The bar and    & & & & & & \\
minicavity    & & & & & & \\
 region        & & & & & & \\
\hline
IRS 2S        & cool                       & $107\pm20$      & yes   & yes   & yes   & yes   \\
IRS 2L        & red                        &  -              & yes   & yes   & yes   & yes\\
IRS 3         & red                        &   -             & yes   & yes   & yes   & yes \\
IRS 12N       & cool                       & $-65\pm5$       & yes   & yes   & yes   & yes \\
IRS 13E       & hot                        & $45\pm60$       & yes   & yes   & yes   & yes\\
IRS 29S       & hot                        & $-93\pm20$      & yes   & no    & yes   & yes\\
              & hot                        & $-164\pm2$      &       &       &       &   \\
\hline
IRS 6W        & hot                        & $63\pm3$        & yes   & yes   & no    & yes\\

\hline
\end{tabular}
\caption{Summary of the observed absorption lines assumed to arise from foreground clouds of the GC sources 
listed in column 1. The spectral types and published radial velocities of the sources are given in columns 2 and 3, respectively. See Table~\ref{tabvel} for references. For IRS~12N we used the average of the values derived by \citet{zhu08} and \citet{figer03}.
}

\end{center}
\end{table*}
\label{tabLOS}

\subsection{The northern arm}
In the spectra of the observed sources in the northern arm (i.e., IRS~5, IRS~10W, and IRS~21), we distinguish the prominent foreground absorption lines at LSR velocities of -40, -60, -85, and -145$\pm$15 km/s and also the three absorptions at -130, -185,~ and +50~km/s. These detections are more specifically observed in the R(0) and P(1) transition line spectra (see Figs.~\ref{IRS5}, \ref{IRS10} and \ref{IRS21}). The -185~km/s absorption is less visible in the R(0) and P(1) spectra of IRS~21 as can be seen in Fig.~\ref{IRS21}.

\subsection{The bar and around the minicavity}\label{bar}
All the observed sources that are located in the bar of the minispiral and around the minicavity, namely IRS~2S, 2L, 12N, 13E, 29S, and 6W (see Fig.~\ref{Slits}), show the four foreground absorptions at -40, -60, -85, and -145 $\pm$15 km/s in their R(0), P(1), and/or P(2) transition spectra (see Figs.~\ref{IRS6W} to \ref{IRS2S}). The spectra also show the 50 km/s absorption in almost all transitions from R(0) to P(3). However, the -185$\pm$15 km/s and -130$\pm$15 km/s clouds are not detected in the spectra of IRS~29S and 6W, respectively.  \\
The IRS~13 spectra in the R(0) and P(1) transitions show an additional line at about $\sim$-210$\pm$15~km/s (see Fig.~\ref{IRS13}). To the south of IRS~13, around the minicavity, the IRS~2L, IRS~2S, and IRS~12N sources also show absorptions at the same velocity (i.e., $\sim$-210$\pm$15 km/s) in their R(0) (except IRS~2S) and P(1) transition spectra (see Figs.~\ref{IRS12N} to ~\ref{IRS2S}). This indicates that the line at this velocity probably arises from material in this IRS13-2L-2S-12N complex around the minicavity. We assume that this material is closer to the GC than the other foreground clouds even though the line is only detected at low transitions, because its detection is confined to only four sources of the central parsec. 

\subsection{The Helium stars}
\label{subsectionHestars}
The  R(0) and P(1) spectra of the helium stars IRS~16C and IRS~16SW show exclusively the lines that we attribute to cold dense LOS clouds with LSR velocities of -40, -60, -85, and -145$\pm$15 km/s as well as the 50$\pm$15 km/s line (see Figs.~\ref{IRS16SW} and \ref{IRS16C}) also at higher transitions. The latter is most probably absorbed in the well-known 50~km/s cloud that is closer to the GC. We note that in the P(1) and the P(2) spectra of IRS~16C, there is a telluric at the position of the $\sim$-40 km/s line and another at the $\sim-145$ km/s line. Moreover, in the R(0) spectrum of IRS~16SW, a telluric is also present close to the $\sim$-40 km/s line. However, these lines are respectively confirmed in the other available spectra of both IRS~16C and IRS~16SW. The helium stars are known not to be embedded in the minispiral. This may explain the absence of lines  other than those absorbed in the four foreground clouds (at -40, -60, -85, and -145 km/s). This also confirms at the same time that the LOS lines are absorbed in a medium located outside the central parsec.\\

In general, most of the individual spectra show nonsaturated \CO12 lines at -40 and -60 km/s, unlike the IRS~1W and IRS~3 spectra. This may indicate a variation of the \CO12 abundance in the corresponding foreground clouds across the field of view and towards the direction of the individual sources of the GC.

\section{Local absorptions in the Galactic center }\label{localabs}
In the last column of Table~\ref{tabvel}, we list the velocities of additional lines that we detect in the spectra of the observed GC sources and that we assume to be associated with the individual sources. These lines are all detected at high transition levels (between P(3) and P(9)) except for IRS~1W, where the detection is only obtained at low levels (R(0) and P(1)) due to a  lack of data at higher levels. As already mentioned, higher transition levels are excited at higher temperatures, implying that the gas at the origin of these absorptions is closer to the sources. In the following, for each of the sources  we enumerate the observed local absorptions and discuss our findings. All results are summarized in Table~\ref{tabvel}.  \\

{\bf IRS1W}: We find absorptions at 35$\pm$15 km/s and  50$\pm$15 km/s in the spectra of the R(0) and P(1) transitions, respectively (see Fig.~\ref{IRS1W}). The first one is consistent with the value of the radial velocity of 35$\pm$20 km/s as derived by \citet{paumard06}. We therefore assume that this absorption could be associated with the local medium of the source even though at this velocity we already concluded that the absorption is also attributed to the 50 km/s cloud. \\

{\bf IRS~2L and IRS~13E}: In the IRS~13E spectra shown in Fig.~\ref{IRS13}, one can clearly distinguish two absorptions at about $\sim$50 and 60$\pm$15 km/s in all available ro-vibrational transition spectra (R(0), P(1), P(2), and P(3)). The 50 km/s line is due to the 50 km/s cloud as stated above and the 60 km/s absorption is consistent with the 45$\pm$60 km/s radial velocity of the source as measured by \citet{genzel00}.\\
However, in all transition spectra of IRS~2L shown in Fig.~\ref{IRS2L}, we also find the two absorptions at around 50 and 60$\pm$15 km/s. Therefore, the 60$\pm$15 km/s may also be coming from the local GC medium in the material of the IRS2L-IRS13 complex. On the other hand, the $\sim$50$\pm$15 km/s is also probably attributed to the 50~km/s GC cloud. Moreover, an additional absorption at -105 km/s in the R(0), P(1), and P(3) spectra of IRS~2L is also detected. It may be a first determination of the proper radial velocity of the source that was not measured before or is possibly related to the minispiral material in the IRS2L-2S-12N region, since this velocity is also observed in the R(0) and P(1) line absorptions of IRS~2S and IRS~12N.\\

{\bf IRS~2S}: A faint absorption at -110 km/s is visible in R(0), P(1), and P(2) transition spectra but not in the P(3) spectrum (see Fig.~\ref{IRS2S}). It is not consistent with the 107$\pm$20 km/s radial velocity measured by \citet{genzel00}. Hence, the absorption is possibly related to the gaseous stream of the
minispiral ISM (see, e.g., the velocity field in \citet{zhao10}) rather than to the emission line of the object IRS~2S as it is also observed in the R(0) and P(1) spectra of IRS~12N and the R(0), P(1), and P(2) spectra of IRS~2L.\\

{\bf IRS~3}: The absorption at 60$\pm$15 km/s is present in the R(0), P(1), and P(2) transition spectra and also in the \CO12 P(6) and P(7) spectra (see Fig.~\ref{IRS3}). There is no measurement of the radial velocity of this source in the literature. Our results therefore provide a first measurement of the radial velocity of the source even though an intervening cloud also absorbs at the same velocity (i.e., $\sim 50\pm$15 km/s) at low transition levels. \\

{\bf IRS~5}: A prominent absorption is observed in all our spectra around 70km/s$\pm$15 km/s (see Fig.~\ref{IRS5}). This value is inside the error bars of the known radial velocity of this source of 110$\pm$60 km/s as given by \citet{tanner05}.\\

{\bf IRS~6W}: Two absorption lines at -20$\pm$ km/s and 65$\pm$ km/s are present in the P(1), P(2), P(4), and P(5) spectra (see Fig.~\ref{IRS6W}). The 65 km/s absorption can be confused with the 50 km/s foreground absorption but such a velocity is consistent with the radial velocity of the source measured from emission lines (\citet{zhu08, herbst93}). The absorption line at about -20~km/s would then be associated with a gaseous component along the LOS, but it is unclear if it is located close to the GC.\\

{\bf IRS~10W}: Towards this continuum source, the absorption features at 110$\pm$15~km/s and 210$\pm$15~km/s are not very prominent in the P(3) transition spectrum (see Fig.~\ref{IRS10}) but are close to the emission line at 80$\pm$60~km/s in the minispiral gaseous flow (\citet{zhao10}) and the 195$\pm$150~km/s line possibly associated with the star (\citet{tanner05}). In addition, our spectra show a faint absorption at -105$\pm$15~km/s that has to be confirmed at higher S/N.\\

{\bf IRS~12N}: This source is the best example we have, showing that we detect $^{12}$CO gas at the velocity of the source and consequently that it is directly associated with the source. Absorption features at -85$\pm$15 km/s are detected in most of the observed transition lines (R(0), P(1), P(3), P(4), P(6), P(7) and P(9)) (see Fig.~\ref{IRS12N}). This velocity is consistent with the radial velocity of the source of about -65$\pm$2km/s combining the values given by \citet{zhu08} and \citet{figer03}. 
Even though this velocity is also rather coincident with the -85 km/s LOS presumably cold cloud we already mentioned, 
it is also probably associated with warmer material probably closer to the source since high transition levels are detected as well. In addition, as mentioned previously in the IRS~2S and 2L paragraphs, an absorption observed at -110$\pm$15~km/s in the R(0) and P(1) spectra is probably associated with the material of the minispiral in the IRS~2L-2S-12N region.\\

{\bf IRS~16C and SW}: The IRS~16 sources (C and SW) are in the northern part of the IRS16 association of bright He~stars. IRS16~C is offset by about 2-3 arcseconds from the bar feature in the minispiral and IRS~16SW just north of or slightly within the bar (see Fig.~\ref{Slits}). As mentioned in Sect.~\ref{subsectionHestars}, we find absorption features in the $^{12}$CO~R(0) transition that we already attributed to LOS clouds. All other higher ro-vibrational transitions we probed do not show any strong absorption line that may be characteristic of local absorption (see Figs.~\ref{IRS16SW} and \ref{IRS16C}).\\

{\bf IRS~21}: An absorption feature at -60$\pm$ 15km/s is detected in the P(3) transition spectrum (see Fig.~\ref{IRS21}) that can be confused with the foreground cloud discussed in Sect.~\ref{sectionLOS}. Within the uncertainties it is also consistent with the radial velocity of the source of -90$\pm$20 km/s as obtained by \citet{genzel00}. Also within the uncertainties, an absorption at 55$\pm$ 15km/s visible in the R(0), P(1), P(2), and P(3) ro-vibrational spectra may be consistent with the 50~km/s cloud.\\

{\bf IRS~29}: The spectra of IRS~29 in the R(0) and P(1) transitions show the LOS absorption at -60 km/s but this absorption is also visible in the P(3) higher-transition spectrum. Since it is compatible with the value of the radial velocity found by \citet{genzel00}, we assume that this absorption is also associated with the source.\\

\newpage

\begin{figure*}
\includegraphics[width=25pc,angle=0]{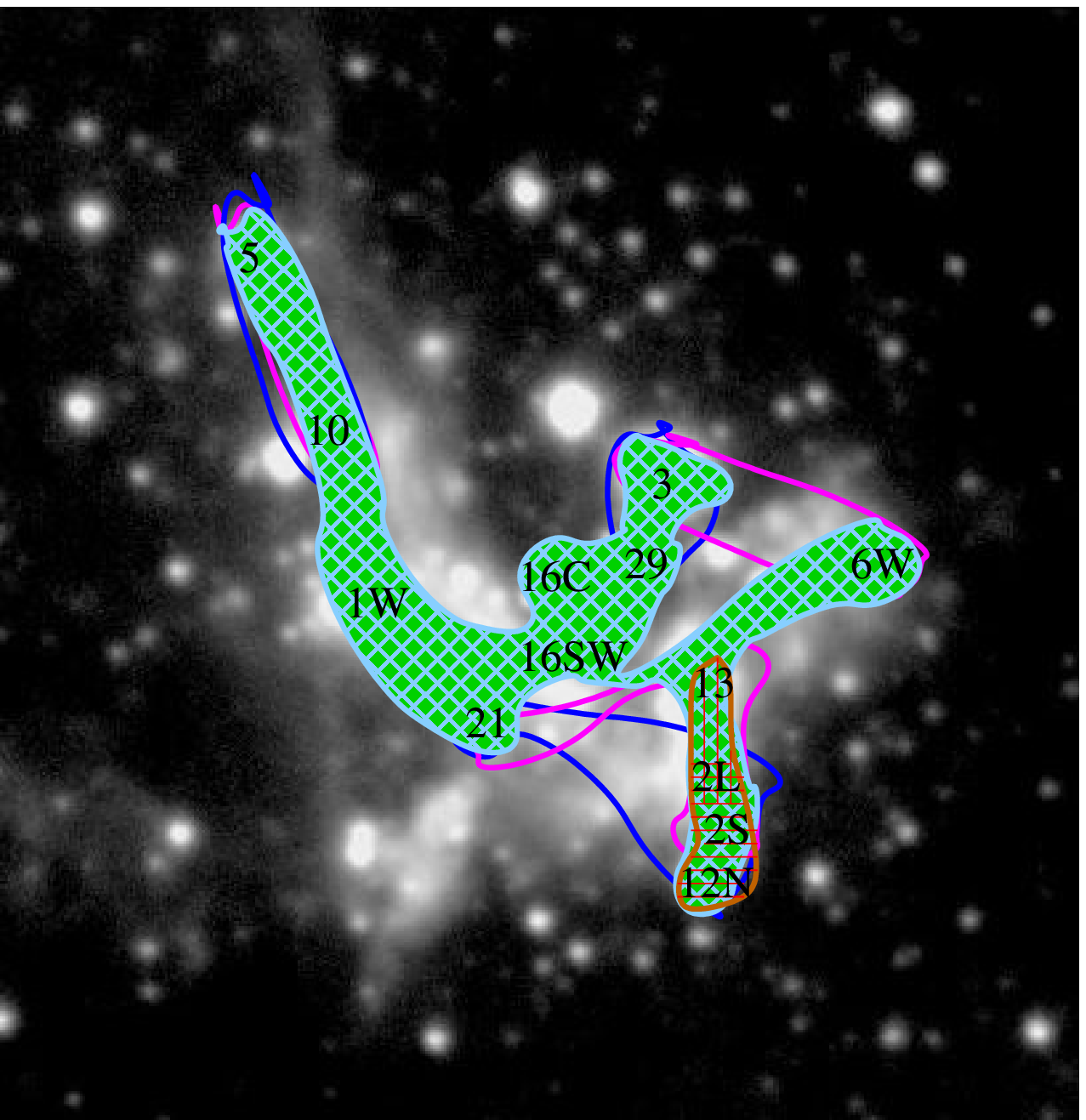} 
\caption{\label{clouds}  
Schematic view of the bright-continuum-emitting regions in the central parsec 
that are affected by absorption of -145, -85, -60, -40 km/s and 50 km/s clouds (in light blue dashed with green), the -185 km/s cloud (in magenta), the -130 km/s cloud (in blue), the -210 km/s IRS13-2L-2S-12N complex cloud (in brown), and the 60 km/s and -110 km/s minispiral material (dashed red vertically and horizontally, respectively).} 
\end{figure*}

\begin{figure*}
\includegraphics[width=25pc,angle=0]{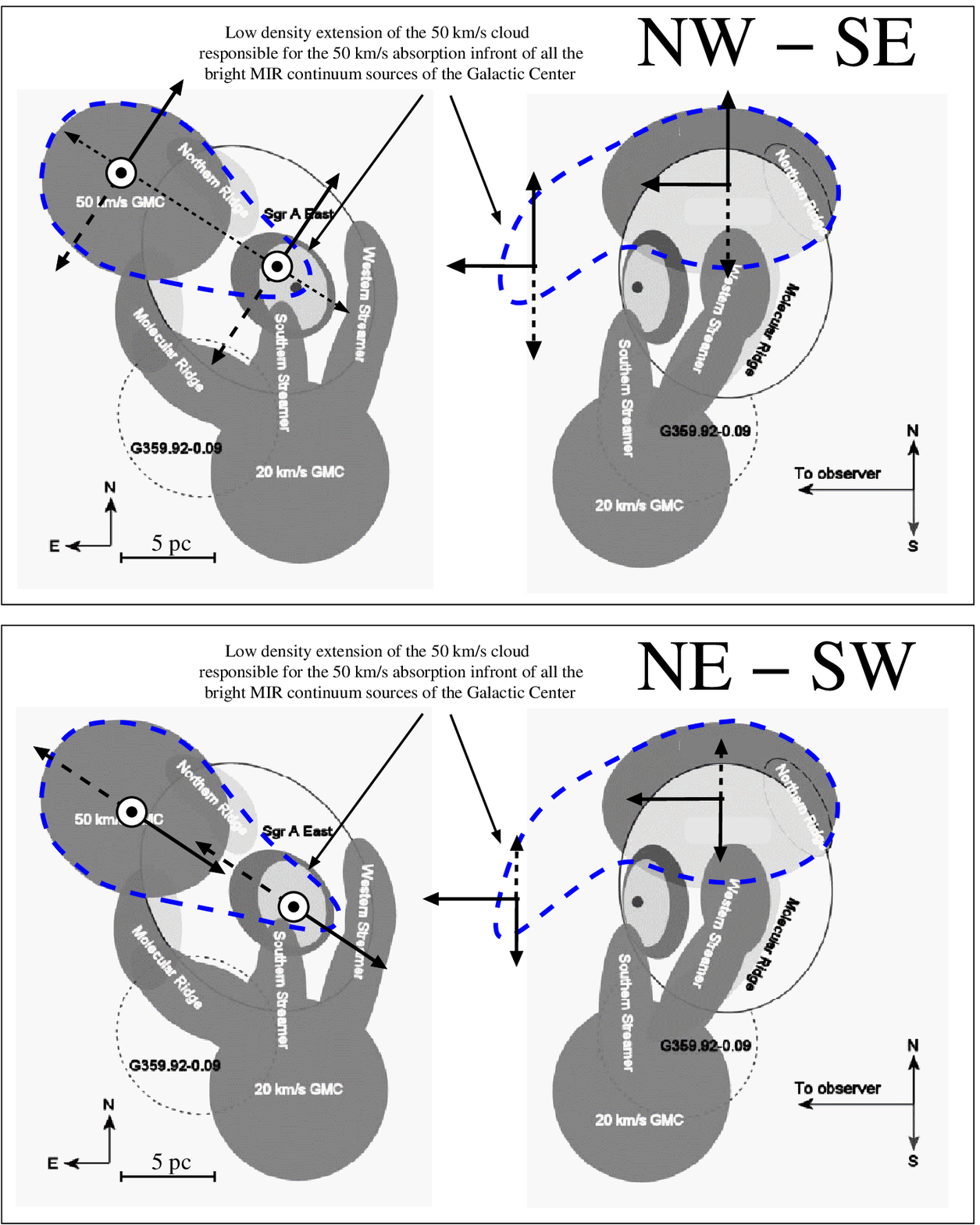} 
\caption{\label{leefig} 
Schematic view of the 50~km/s cloud. It is one of the closest cloud complexes to the GC.
Following \citet{lee08} we show the relative locations of some of the clouds
that are quite close to the LOS. The circumference of the 50~km/s cloud and its possible extension have been indicated in light blue.
In the right panels, the radial velocity vector is shown as a narrow peak pointing towards the observer.
In the left panels, the radial velocity vector is shown as a solid horizontal 
vector pointing towards the observer. The approximate linear scale is indicated.
The {\bf top panel} shows the situation if the 50~km/s cloud and its extension 
have proper motions towards the NW (bold solid line) or SE (bold dashed line). 
We have drawn the radial and proper motion vectors (and their projections) 
in both top panel views. 
The direction perpendicular to the orbital plane of the bypassing 50~km/s cloud
complex is shown as a short dashed line.
The {\bf bottom panel} shows the situation for a proper motion in NE (bold solid line) 
or SW (bold dashed line).}
\end{figure*}

\section{Conclusion and discussion}
In this paper we present the results derived from our high-spectral-resolution CRIRES observations of 13 IR sources located in the central parsec of the Galaxy. The data were obtained around the \CO12 R(0), P(1), P(2), P(3), P(4), P(5), P(6), P(7), and P(9) transition lines and provide the first evidence of a complex structure in the ISM along the LOS and in the close environment of the central sources. We have found evidence for four foreground cold clouds at radial velocities v$_{LSR}$ of the order of -145, -85, -60, and -40$\pm$15 km/s that show absorptions in the lower transition levels from R(0) to P(2) and in all the observed spectra. We also found evidence for two other cold clouds at radial velocities v$_{LSR}$ of the order of -185 and -130$\pm$15 km/s that seem to be closer to the GC since the absorption lines at the velocities are not present in all the spectra. Specifically, these lines are absent in the spectra of the helium IRS~16 sources that are away from  the minispiral structure, indicating that the corresponding cloud may be associated with the minispiral material with a temperature around 300~K (e.g., \citet{garciamarin11}, \citet{kunn12}, \citet{serabyn86}).\\
 The same argument leads to the conclusion that a closer cloud at a  velocity of -210$\pm$15 km/s, and maybe another one also at 60$\pm$15 km/s, is probably present in front of the IRS~13, IRS~2L complex as the spectra of these sources show absorption lines at these velocities. Moreover, in all sources we find an absorption in velocity range of 50-60 km/s, possibly associated with the 50~km/s cloud in the GC region. This opens up the possibility that the 50 km/s cloud has a low-density extension in front of the GC giving rise to the observed absorptions (see detailed discussion on the 50~km/s cloud extension further in the conclusion). This provides us with a clearer picture of the distribution of the different intervening clouds. In Fig.~\ref{clouds}, we propose a rough projection of the distribution of these clouds in the field of view. \\

We also detected additional absorption lines in the spectra of most of the individual sources at high transition levels (from P(3) to P(9)). For most of them, the measured velocities fit the radial velocities of the sources themselves as obtained from the literature. This indicates that the absorptions are taking place in the very close environment of the sources at the temperatures deduced from these transition lines of a few tens of Kelvin up to 300~K (e.g., \citet{mon01}, \citet{garciamarin11}, \citet{kunn12}, \citet{serabyn86}). At the same time, we were able to determine proper radial velocities of IRS~2L and IRS~13N that have not been measured previously.\\
The fact that for a number of sources the low R and P ro-vibrational absorption lines are also at the radial velocities of the sources implies that the absorptions either take place in colder LOS clouds (which is less likely) or it may provide further evidence for the fact that low-temperature material can in fact survive in clouds during the short (in comparison to the evaporation timescale) transient time through the central cluster as we claimed in our previous works (e.g., \citet{moultaka15a, moultaka15b}).\\

{\bf The proposed  extension of the 50~km/s cloud:}

As mentioned above, since all of the continuum sources in the central parsec show an absorption at 50~km/s, this may point toward the existence of an extension of the 50~km/s cloud which must be floating by the GC northeast of the center (see Figs.~\ref{LOS} and \ref{leefig} presenting a possible 3D distribution of this extension along the LOS towards the GC). Alternatively, the absorption towards the central continuum sources shows up incidentally at approximately the same velocity as the 50~km/s cloud. 

The presence of an H$_2$CO absorption in front of Sgr~A East but not Sgr~A West lead to the conclusion that the 50 km/s cloud is located behind Sgr~A West (\citet{whiteoak74}). 
This finding has been supported by several other investigations as was pointed out by \citet{geballe89} and \citet{goto14}. The abundance ratio between CO and H$_2$CO is of the order of $10^5$~(\footnote{With respect to molecular hydrogen, \citet{pauls96} give a Formaldehyde abundance towards Sgr A of a few times $10^{-10}$.}). The ratio of the absorption line strengths between Sgr~A East and West is only of the order of 12(\footnote{Whiteoak, Rogstad \& Lockhart (1974) find H$_2$CO optical depths toward Sgr~A East  2.5 and less than 0.03 toward Sgr~A West in a 20''$\times$40'' beam.}). Hence, in a case of low to moderate optical depth, the presence of CO absorption in the absence of a H$_2$CO absorption in front of Sgr~A West is relatively conceivable.

However, \citet{geballe89} already detected CO absorption features in the spectra of IRS~3 and IRS~7 speculating that these absorptions may be due to the 50~km/s cloud. At that point the conclusion was unequivocal as the velocity of the western edge of the CND is also close to 50~km/s. Also, \citet{goto14} suggested that the 60 km/s absorption observed in their spectra is probably associated with the CND. Moreover, \citet{geballe89} did not detect 50~km/s absorption towards IRS~1 and IRS~2. Our detection of absorption close to 50~km/s towards most continuum sources implies that a 50~km/s cloud filament indeed lies in front of the central stellar cluster. 
If the absorptions were predominantly from the CND then in the eastern region of the stellar cluster we would expect them to occur
rather at -20 to -50 km/s (e.g., \citet{Amo10}).
Hence, we are forced to conclude that a 50km/s cloud is in front of the entire central stellar cluster and may be associated with the 50 km/s cloud whose bulk however is located behind Sgr~A West.
\\
One may question how a dynamically stabile configuration can be explained that allows for an extension of the 50~km/s cloud into the LOS towards the continuum sources in the GC close to Sgr~A. To answer this question, we can compare the radial motion of 50~km/s with the expected orbital velocity at the distance of about 5~pc to the GC implied by \citet{lee08}. Using the enclosed mass plot given by \citet{schoedel03}, we find an enclosed mass of $M_6 = M/(10^6 {M}_{\odot}) \sim 10$. With the relation $v[km/s]=2.07 \sqrt{M_6/r[kpc]}$, this results in an expected orbital velocity of about 92 km/s. This implies that the proper motion of the 50~km/s cloud is at least 77~km/s or higher. Hence, the entire 50 km/s cloud including its extension suggested by our absorption line profiles may be moving, for example, towards the NW or SE, passing by the GC. This situation is shown in the top two panels of Fig.~\ref{leefig}.\\
As shown in the bottom two panels of Fig.~\ref{leefig}, a proper motion in the NE or SW direction is less likely. If the 50~km/s cloud was on a NE trajectory (dashed line in the bottom panel of Fig.~\ref{leefig}) it would come almost straight from the central cluster region. A survival of that passage or the presence of the nevertheless rather fragile and transient CND (\citet{reqtor12})
would be very unlikely. Flying towards the SW would imply that the 50~km/s cloud is heading towards the central cluster region (bold line in the bottom panel of Fig.~\ref{leefig}). In both cases one would expect a significant velocity gradient and a velocity difference between the main cloud and its extension - also in the radial direction - but this is not the case.\\
However, the flight directions NW and SE could be realized as a periapse passage of the 50~km/s cloud. In this case, the main cloud and its extension could be passing by the central region 
in one common orbital plane at approximately the same distance and argument of periapsis (top panels in Fig~\ref{leefig}). The radial velocities would be very similar and differences between the 50~km/s cloud and its extension would mainly be dominated by a relatively small velocity gradient perpendicular to the orbital plane of this 
cloud system. In this scenario, part of the cloud could be in  front of the central cluster region with the same radial velocity as the bulk of the cloud.\\

A very good candidate for this 50 km/s cloud section in
front of the central stellar cluster is the central association
(CA) found in 50 km/s molecular line emission with ALMA
(\citet{moser17}).
 This cloud complex is most likely identical
to the 50 km/s OH absorption component found by \citet{karlsson15}.
In both cases, the gas covers the central region with an extent
of about 20 arcseconds including Sgr~A* and the central  stellar
cluster.
The arrangement of OH-absorbing gas at 59 km/s shown by \citet{karlsson15}
in their Fig.~1 (see also Fig.~12 by  \citet{moser17})
is very similar to our proposed extent of the 50km/s cloud
(blue dashed contour in the  top left panel of Fig.~\ref{leefig}).
This is particularly true if one considers the fact 
that the component in front
of the central stellar cluster and the 50~km/s cloud region 
are linked at velocities between 20 and 50~km/s as shown in the
OH absorption channel maps by \citet{karlsson15} in their Fig.~2.

In this paper, we have shown that the continuum sources within the central parsec are ideally suited to studying the physical conditions and the location of the absorbing clouds. For the mostly extended dusty sources, future observations with higher angular resolution and point-source sensitivity will potentially allow us to spatially discriminate between absorption against the compact stellar component and the more extended part of the structure that might partially come from the mini-spiral gas flow. Higher sensitivity will also allow for the absorption towards fainter sources to be mapped out such that the spatial distribution of the absorbing gas clouds can be determined in a more complete way. Right now we can only comment on the absorption towards relatively bright sources. The absorbing clouds may however have a larger extent and the 50~km/s absorption may actually cover the entire field (see Fig.~\ref{leefig}). Suitable instruments for future point-source-sensitive observations are MATISSE at the VLT (\citet{lopez18}) and METIS at the EELT (\citet{brandl18}). Instruments that permit further studies of the absorption towards more extended source components are VISIR at the VLT (\citet{lagage03}) and MIRI aboard the JWST (\citet{garciamarin18,wright10}).



\begin{appendix}

\section{Spectra of the Galactic center IR sources}
\begin{figure}[!h]
\begin{minipage}{40pc}
\centering
\includegraphics[width=30pc,angle=0]{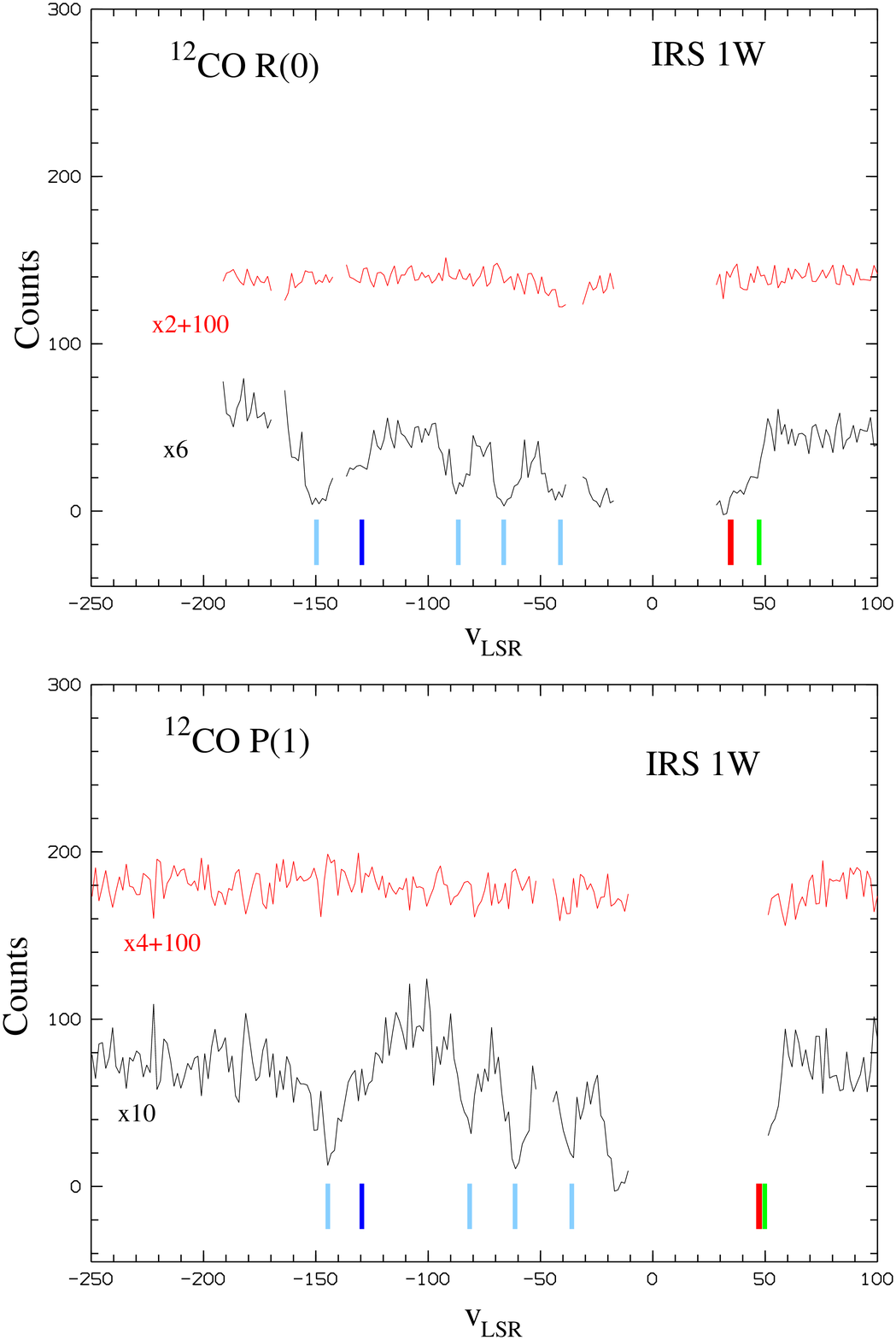}   
\caption{\label{IRS1W} Spectra (in black) of IRS 1W in the \CO12 R(0) and P(1) line transitions. Also shown (in red) are  the spectra of the standard stars observed close to the science observation to correct for telluric lines. The blanked areas correspond to regions in the spectrum where the telluric corrections were not successful. The red arrows indicate the local absorptions at velocities listed in Table~\ref{tabvel}. The other colored arrows indicate the absorptions at the same velocities as in Figs.~\ref{sumR0_P1_P2} and \ref{sumR0P1P2}. The scaling and/or shifting factors of the spectra are given in red and black for the standard star and the GC object, respectively.} 
\end{minipage} 
\end{figure}

\begin{figure*}
\begin{minipage}{40pc}
\includegraphics[width=40pc,angle=0]{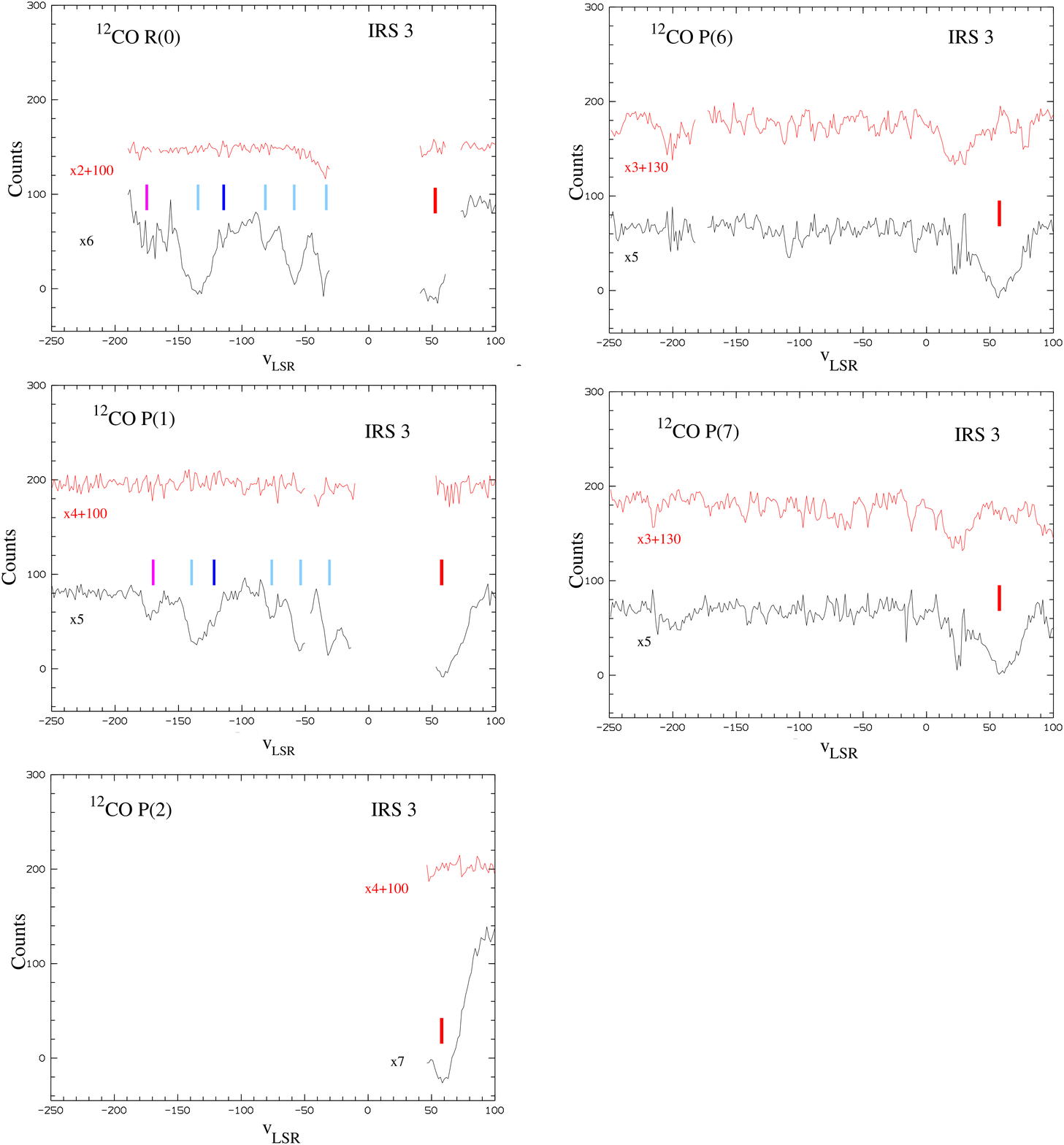} 
\caption{\label{IRS3} Spectra (in black) of IRS 3 in the \CO12 R(0), P(1), P(2), P(6) and P(7) transitions. Also shown (in red) are the spectra of the standard stars observed close to the science observation to correct for telluric lines. The blanked areas correspond to regions in the spectrum where the telluric corrections were not successful. The red arrows indicate the local absorptions at velocities listed in Table~\ref{tabvel}. The other colored arrows indicate the absorptions at the same velocities as in Figs.~\ref{sumR0_P1_P2} and \ref{sumR0P1P2}. The scaling and/or shifting factors of the spectra are given in red and black for the standard star and the GC object, respectively. } 
\end{minipage} 
\end{figure*}

\begin{figure*}
\begin{minipage}{40pc}
\includegraphics[width=40pc,angle=0]{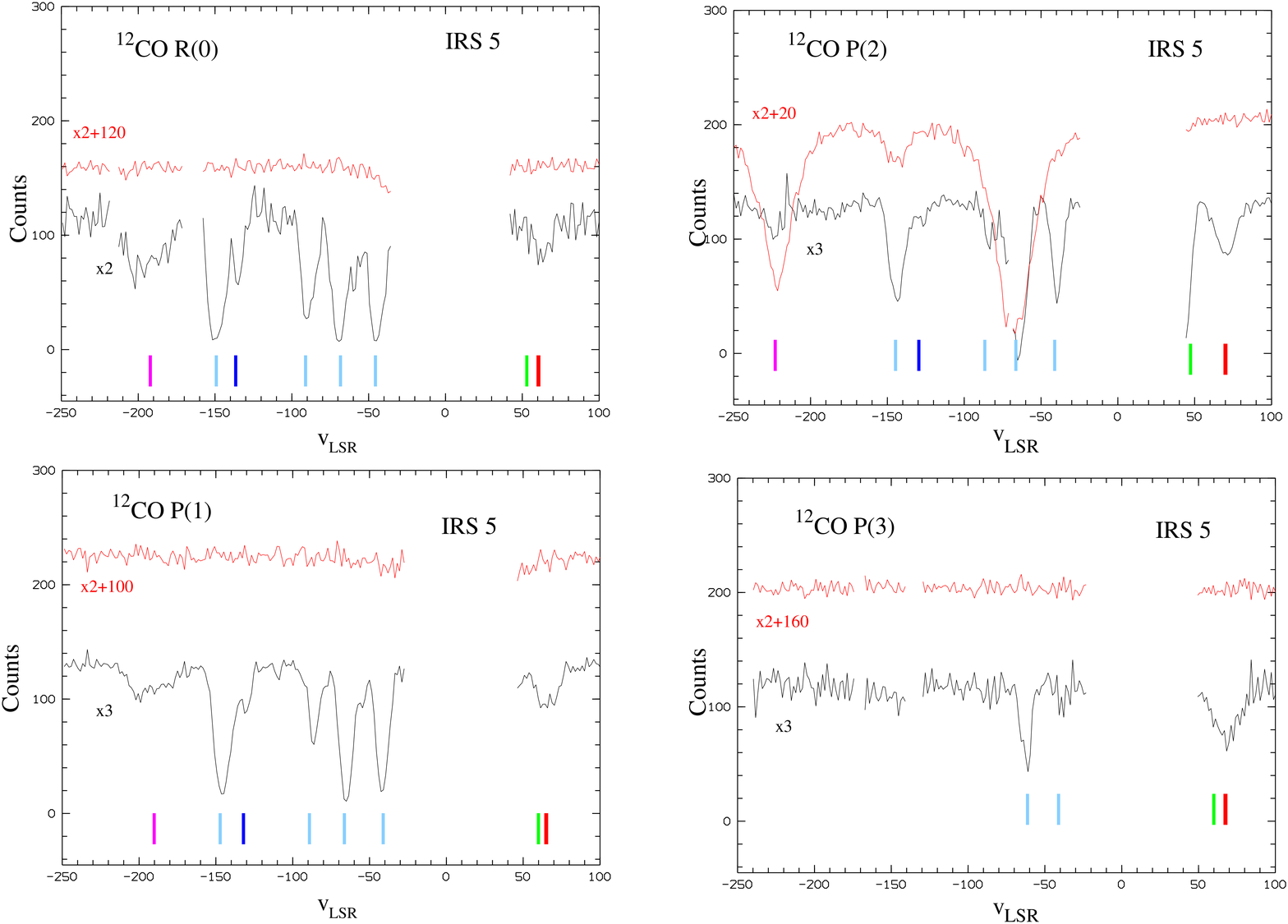}   
\caption{\label{IRS5} Spectra (in black) of IRS 5 in the \CO12 R(0), P(1), P(2) and P(3) transitions. Also shown (in red) are  the spectra of the standard stars observed close to the science observation to correct for telluric lines. The blanked areas correspond to regions in the spectrum where the telluric corrections were not successful. The red arrows indicate the local absorptions at velocities listed in Table~\ref{tabvel}. The other colored arrows indicate the absorptions at the same velocities as in Figs.~\ref{sumR0_P1_P2} and \ref{sumR0P1P2}. The scaling and/or shifting factors of the spectra are given in red and black for the standard star and the GC object, respectively.} 
\end{minipage} 
\end{figure*}

\begin{figure*}
\begin{minipage}{40pc}
\includegraphics[width=40pc,angle=0]{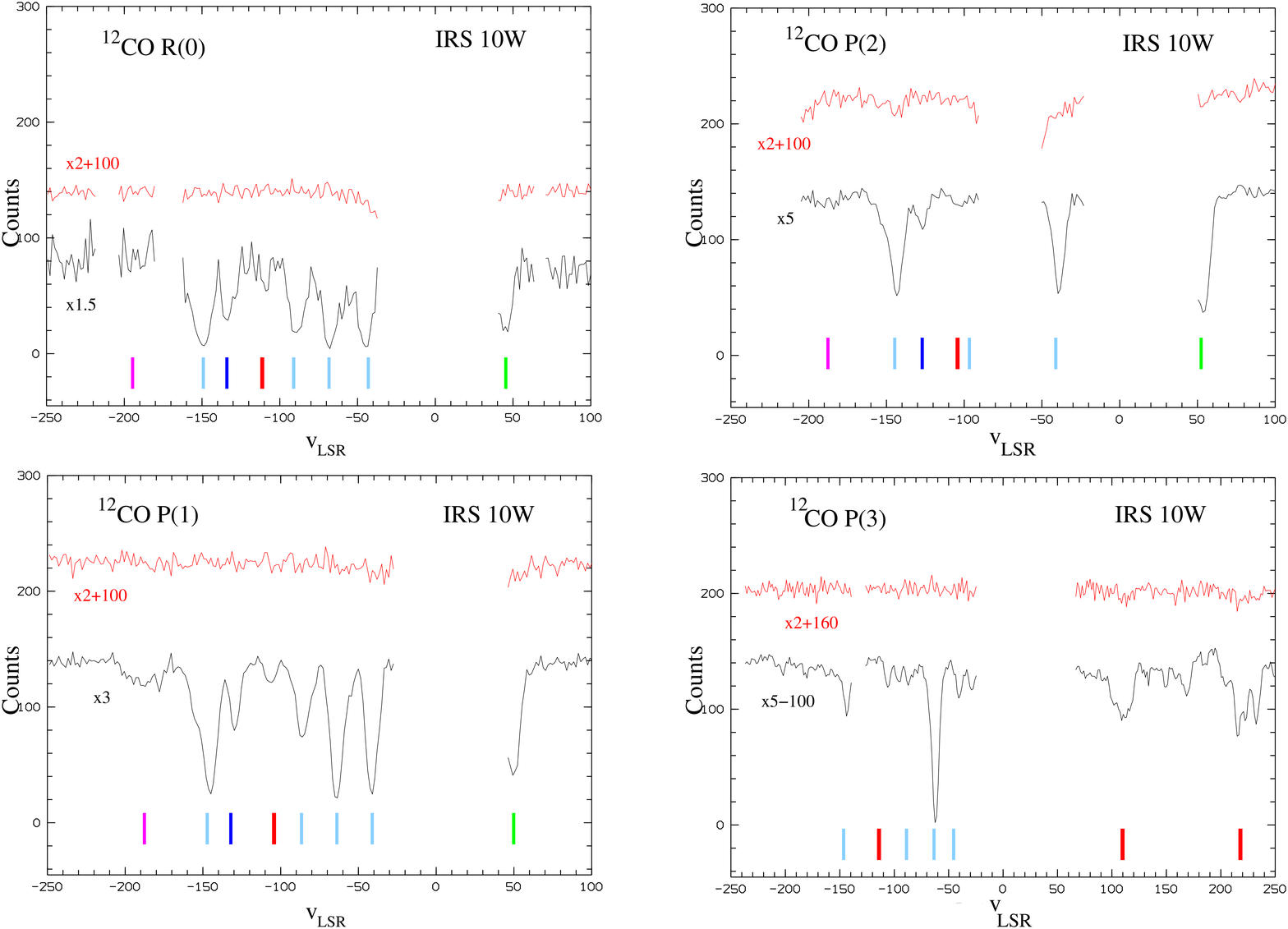} 
\caption{\label{IRS10} Spectra (in black) of IRS 10W in the \CO12 R(0), P(1), P(2) and P(3) transitions. Also shown (in red) are  the spectra of the standard stars observed close to the science observation to correct for telluric lines. The blanked areas correspond to regions in the spectrum where the telluric corrections were not successful. The red arrows indicate the local absorptions at velocities listed in Table~\ref{tabvel}. The other colored arrows indicate the absorptions at the same velocities as in Figs.~\ref{sumR0_P1_P2} and \ref{sumR0P1P2}. The scaling and/or shifting factors of the spectra are given in red and black for the standard star and the GC object, respectively.} 
\end{minipage} 
\end{figure*}

\begin{figure*}
\begin{minipage}{40pc}
\includegraphics[width=40pc,angle=0]{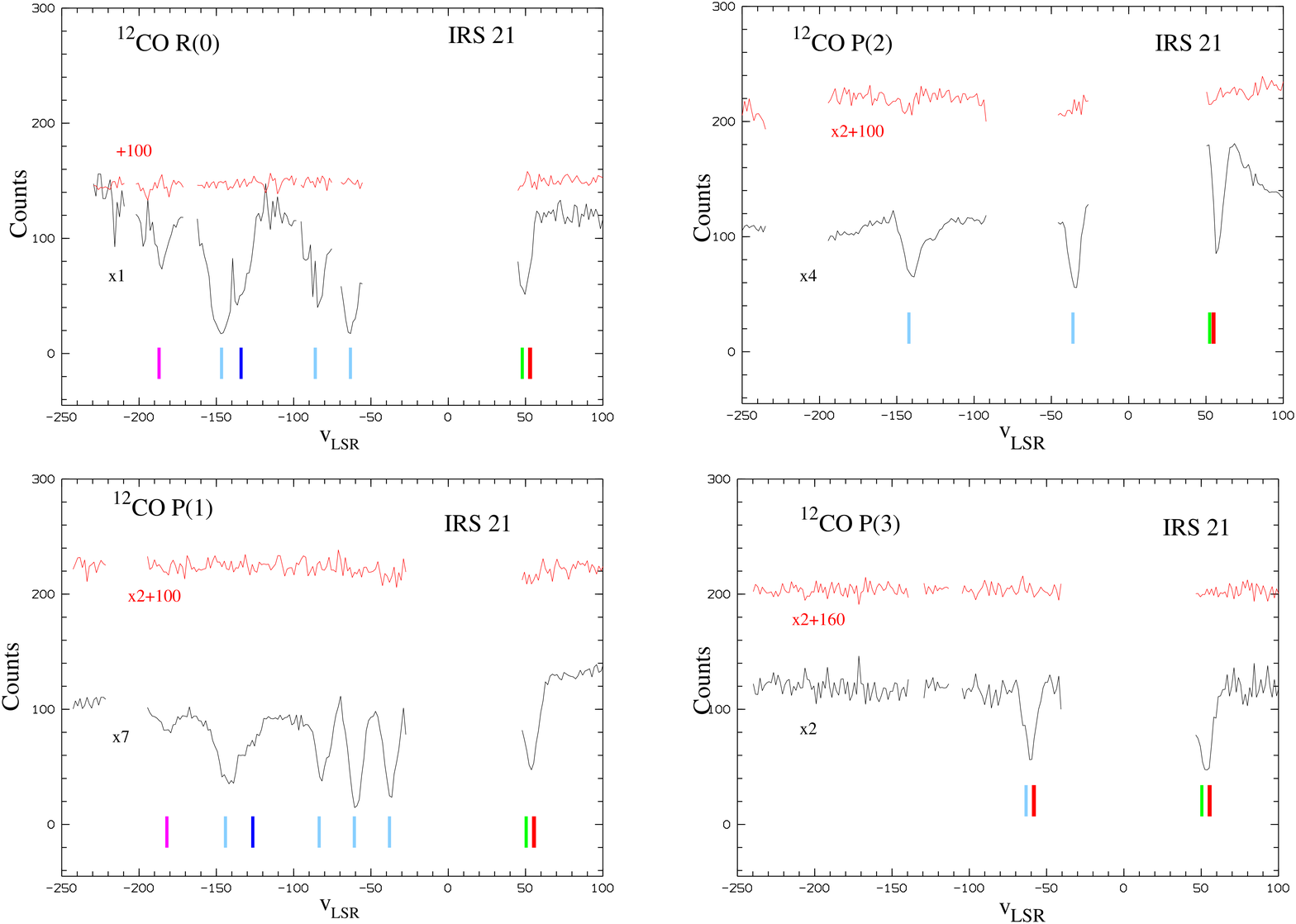} 
\caption{\label{IRS21} Spectra (in black) of IRS 21 in the \CO12 R(0), P(1), P(2) and P(3) transitions. Also shown (in red) are  the spectra of the standard stars observed close to the science observation to correct for telluric lines. The blanked areas correspond to regions in the spectrum where the telluric corrections were not successful. The red arrows indicate the local absorptions at velocities listed in Table~\ref{tabvel}. The other colored arrows indicate the absorptions at the same velocities as in Figs.~\ref{sumR0_P1_P2} and \ref{sumR0P1P2}. The scaling and/or shifting factors of the spectra are given in red and black for the standard star and the GC object, respectively.}  
\end{minipage}
\end{figure*}

\begin{figure*}
\begin{minipage}{40pc}
\includegraphics[width=40pc,angle=0]{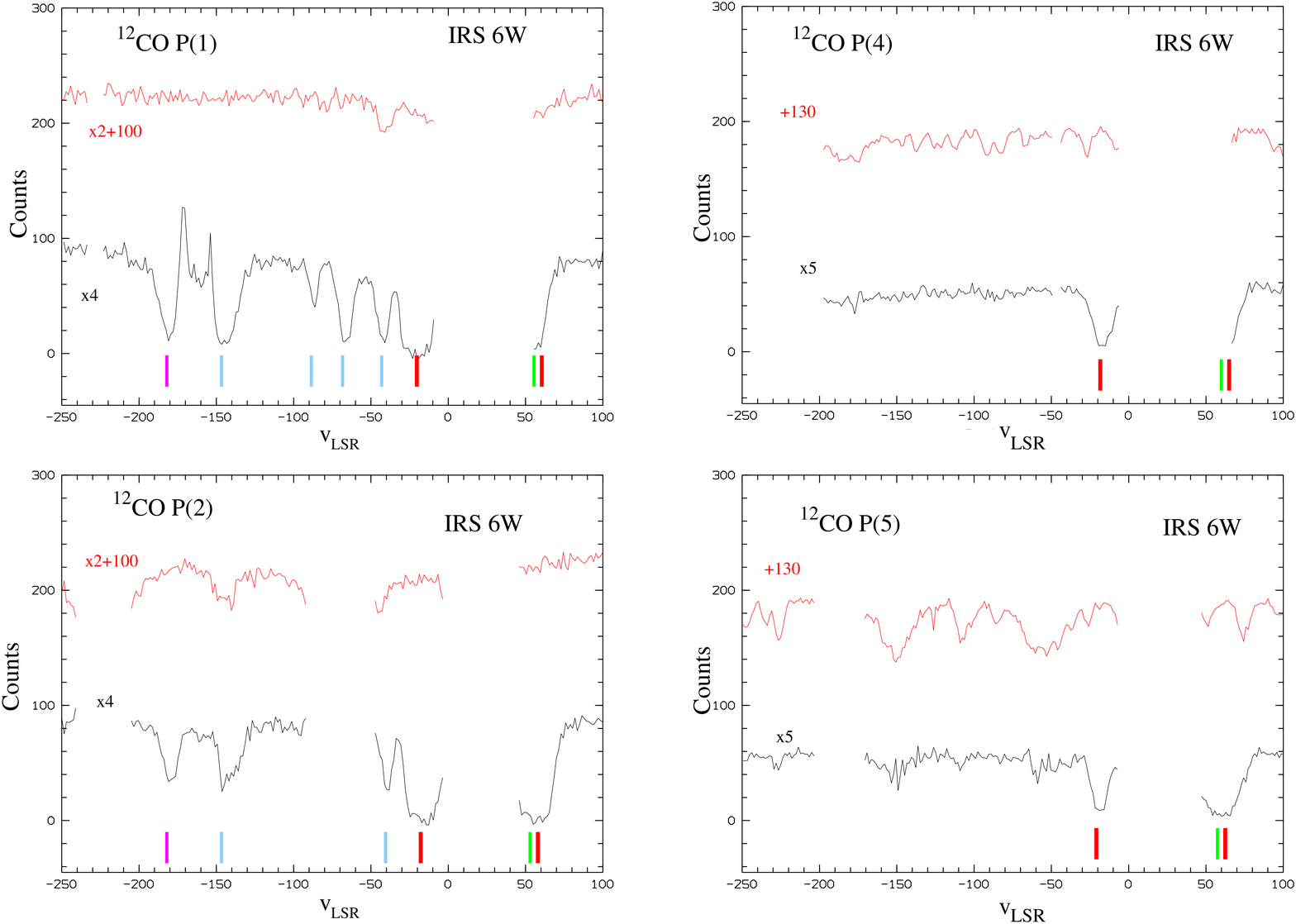} 
\caption{\label{IRS6W} Spectra (in black) of IRS 6W in the \CO12 P(1), P(2), P(4) and P(5) transitions. Also shown (in red) are  the spectra of the standard stars observed close to the science observation to correct for telluric lines. The blanked areas correspond to regions in the spectrum where the telluric corrections were not successful. The red arrows indicate the local absorptions at velocities listed in Table~\ref{tabvel}. The other colored arrows indicate the absorptions at the same velocities as in Figs.~\ref{sumR0_P1_P2} and \ref{sumR0P1P2}. The scaling and/or shifting factors of the spectra are given in red and black for the standard star and the GC object, respectively. } 
\end{minipage}
\end{figure*}

\begin{figure*}
\begin{minipage}{40pc}
\includegraphics[width=40pc,angle=0]{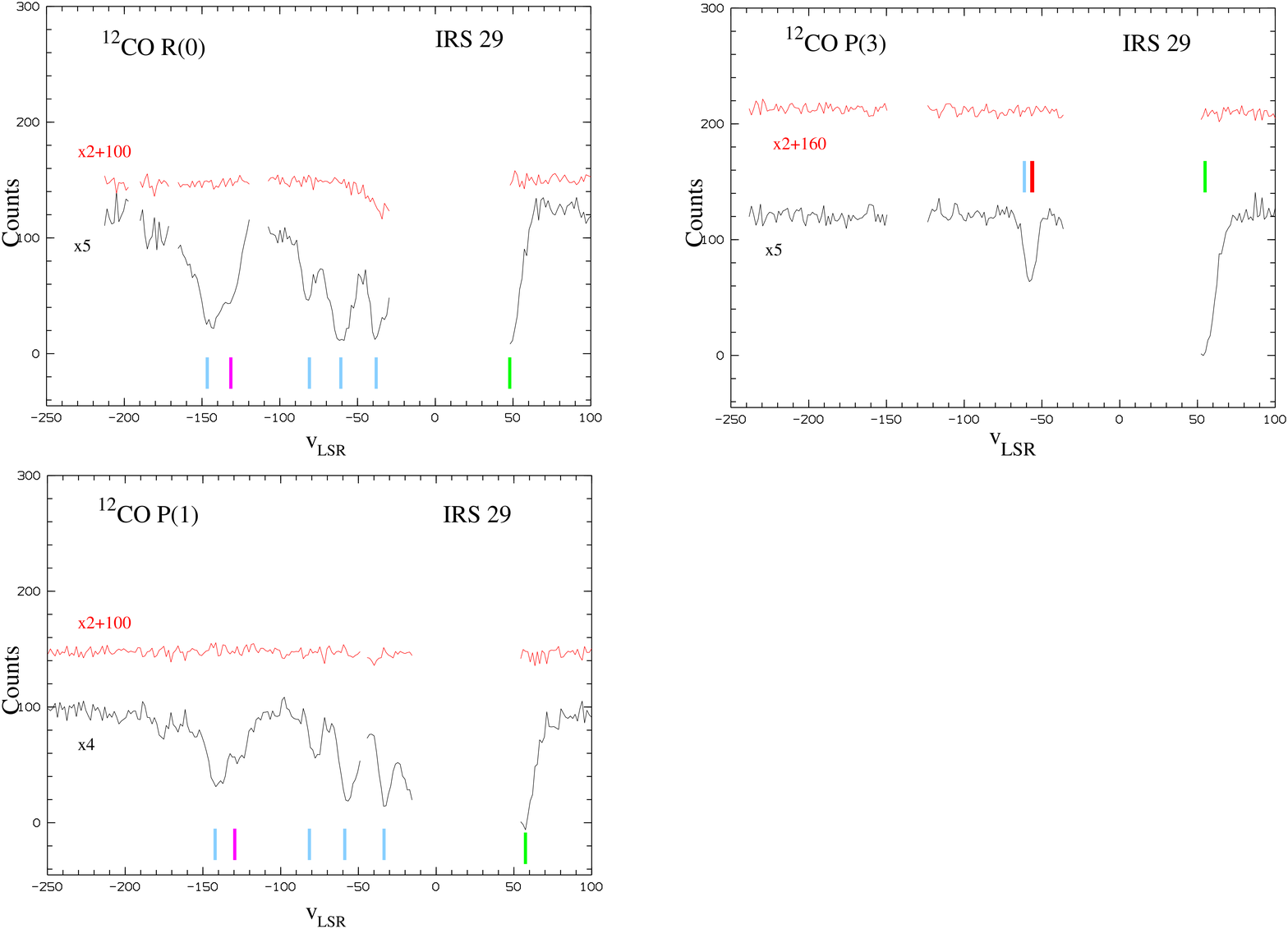} 
\caption{\label{IRS29} Spectra (in black) of IRS 29 in the \CO12 R(0), P(1) and P(3) transitions. Also shown (in red) are the spectra of the standard stars observed close to the science observation to correct for telluric lines. The blanked areas correspond to regions in the spectrum where the telluric corrections were not successful. The red arrows indicate the local absorptions at velocities listed in Table~\ref{tabvel}. The other colored arrows indicate the absorptions at the same velocities as in Figs.~\ref{sumR0_P1_P2} and \ref{sumR0P1P2}. The scaling and/or shifting factors of the spectra are given in red and black for the standard star and the GC object, respectively.} 
\end{minipage}
\end{figure*}

\begin{figure*}
\begin{minipage}{40pc}
\includegraphics[width=40pc,angle=0]{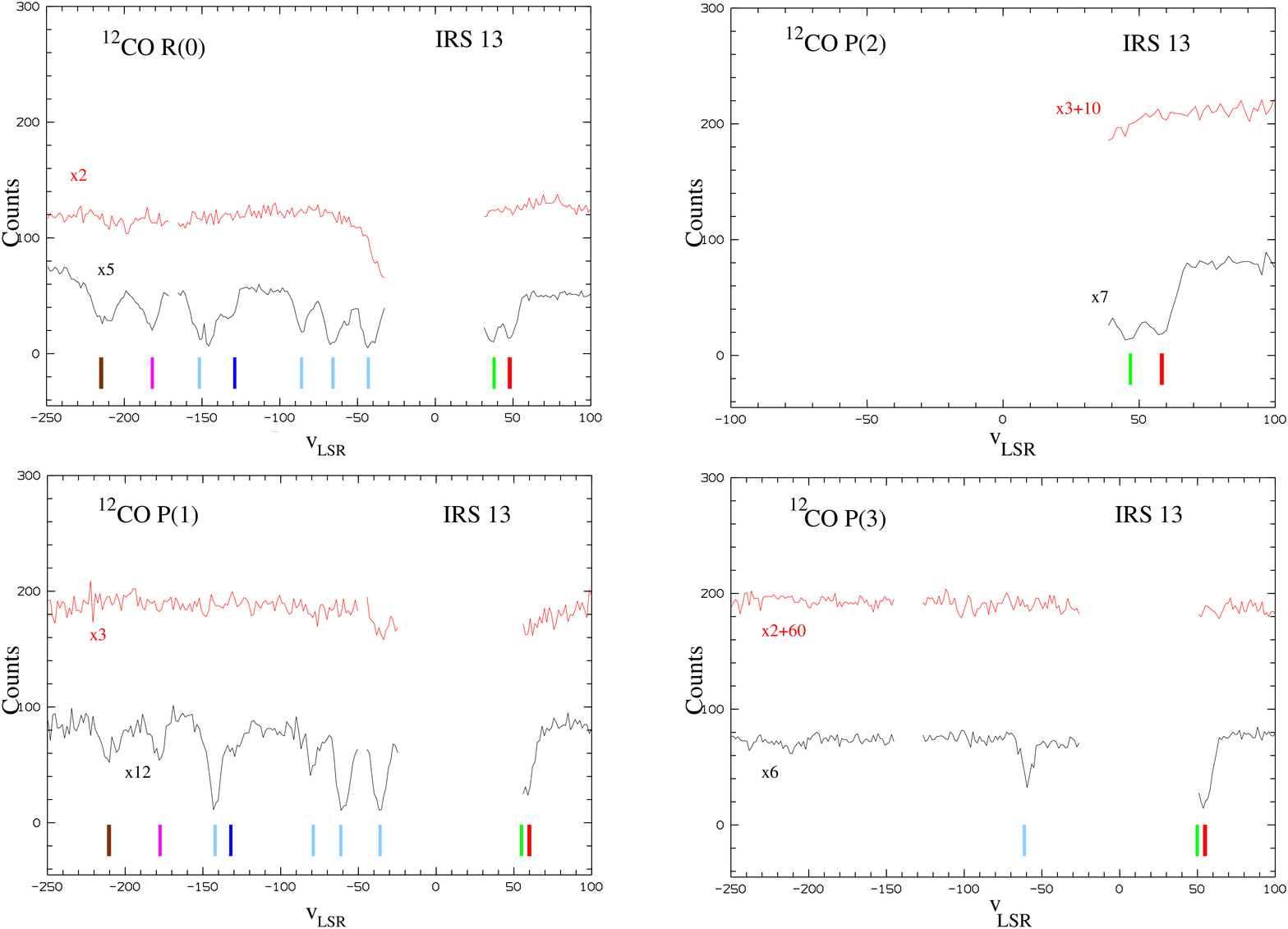} 
\caption{\label{IRS13} Spectra (in black) of IRS 13 in the \CO12 R(0), P(1), P(2) and P(3) transitions. Also shown (in red) are  the spectra of the standard stars observed close to the science observation to correct for telluric lines. The blanked areas correspond to regions in the spectrum where the telluric corrections were not successful. The red arrows indicate the local absorptions at velocities listed in Table~\ref{tabvel}. The other colored arrows indicate the absorptions at the same velocities as in Figs.~\ref{sumR0_P1_P2} and \ref{sumR0P1P2}. Finally, the brown arrows indicate the line at -210~km/s. The scaling and/or shifting factors of the spectra are given in red and black for the standard star and the GC object, respectively.}  
\end{minipage}
\end{figure*}

\begin{figure*}
\begin{minipage}{40pc}
\includegraphics[width=40pc,angle=0]{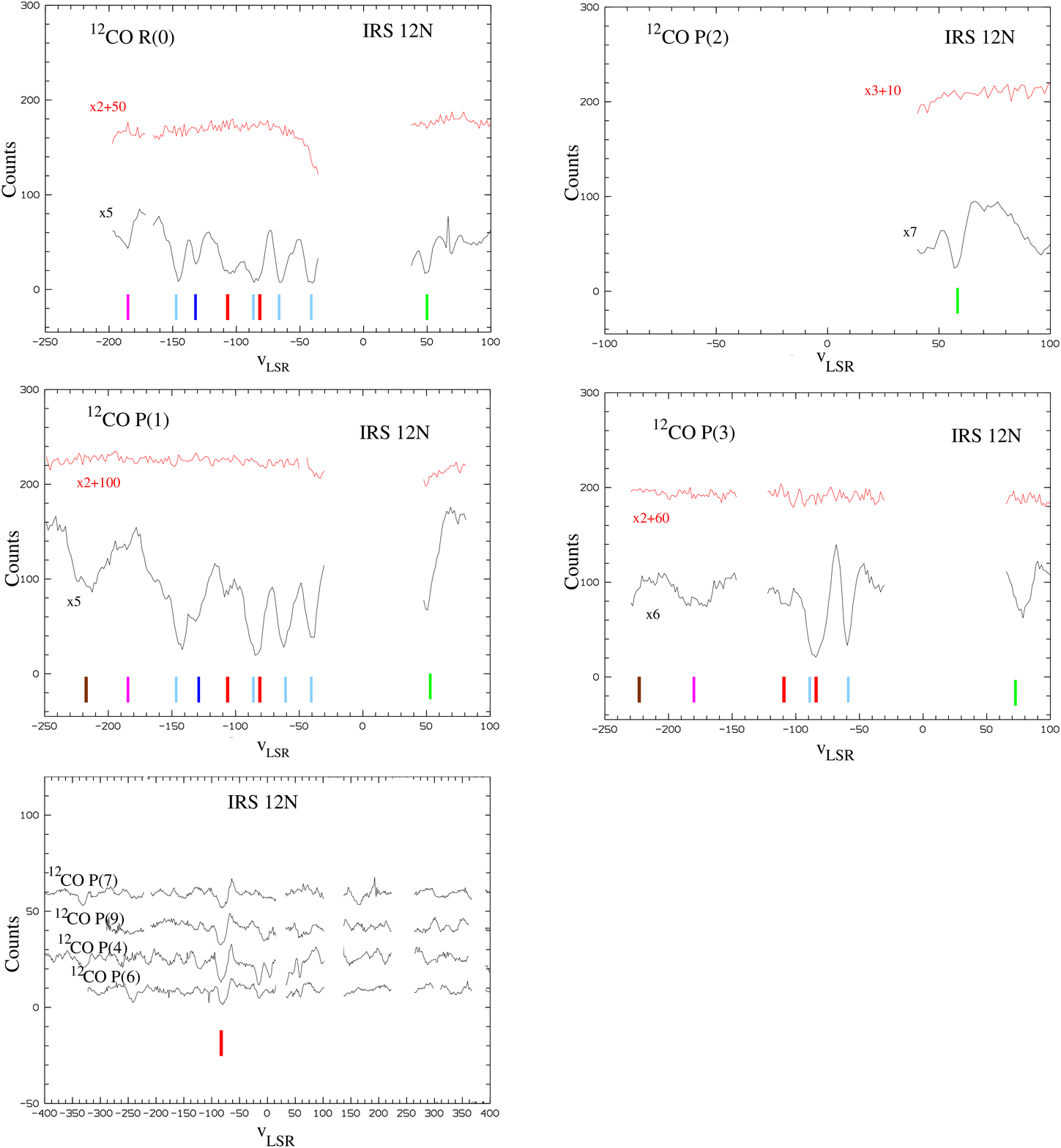} 
\caption{\label{IRS12N} Spectra (in black) of IRS 12N in the \CO12 R(0), P(1), P(2), P(3), P(4), P(6), P(7) and P(9) transitions. Also shown (in red) are  the spectra of the standard stars observed close to the science observation to correct for telluric lines. The blanked areas correspond to regions in the spectrum where the telluric corrections were not successful. The red arrows indicate the local absorptions at velocities listed in Table~\ref{tabvel}. The other colored arrows indicate the absorptions at the same velocities as in Figs.~\ref{sumR0_P1_P2} and \ref{sumR0P1P2}. Finally, the brown arrows indicate the line at -210~km/s. The scaling and/or shifting factors of the spectra are given in red and black for the standard star and the GC object, respectively.}  
\end{minipage}
\end{figure*}

\begin{figure*}
\begin{minipage}{40pc}
\includegraphics[width=40pc,angle=0]{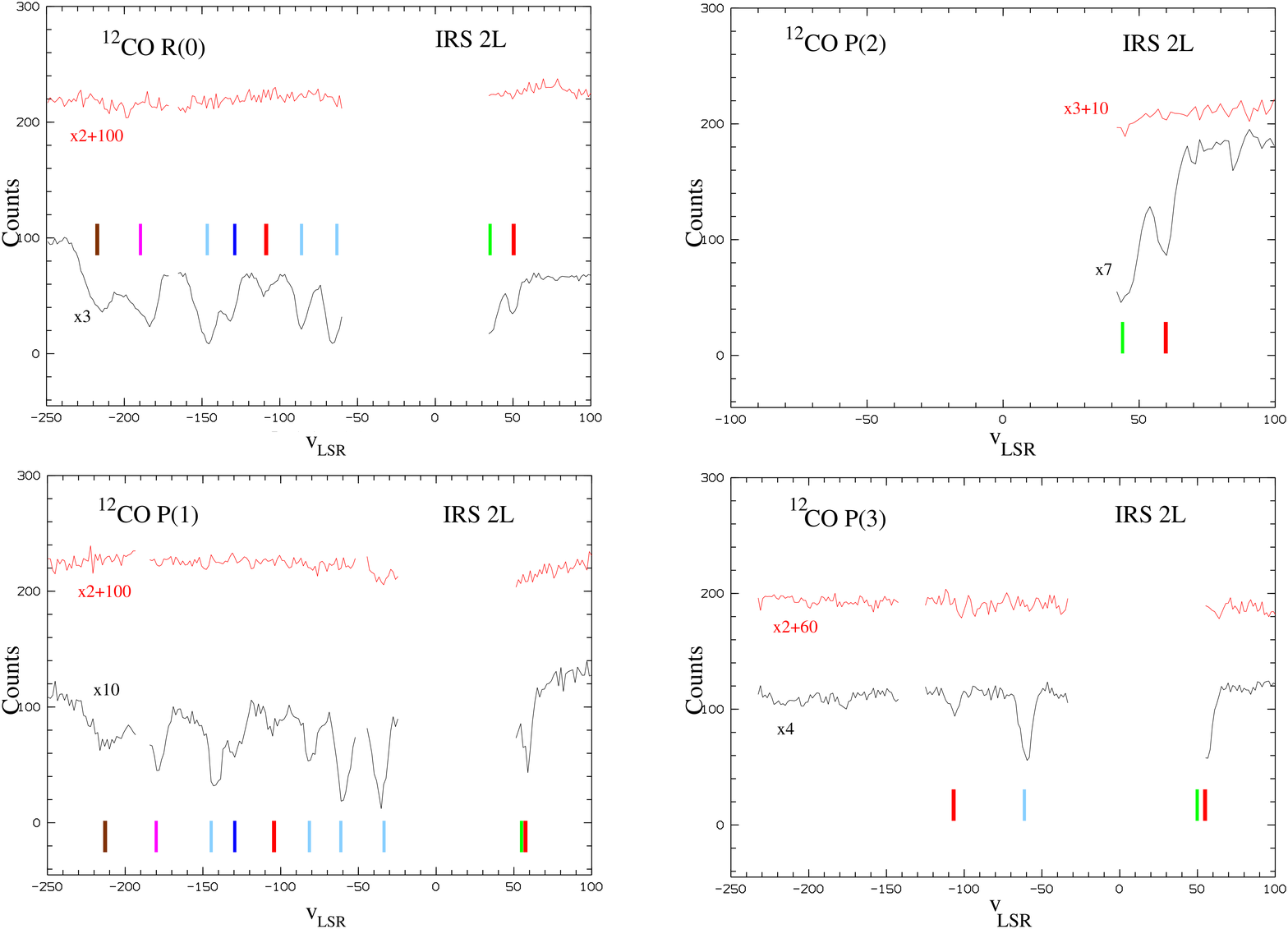} 
\caption{\label{IRS2L} Spectra (in black) of IRS 2L in the \CO12 R(0), P(1), P(2) and P(3) transitions. Also shown (in red) are  the spectra of the standard stars observed close to the science observation to correct for telluric lines. The blanked areas correspond to regions in the spectrum where the telluric corrections were not successful. The red arrows indicate the local absorptions at velocities listed in Table~\ref{tabvel}. The other colored arrows indicate the absorptions at the same velocities as in Figs.~\ref{sumR0_P1_P2} and \ref{sumR0P1P2}. Finally, the brown arrows indicate the line at -210~km/s. The scaling and/or shifting factors of the spectra are given in red and black for the standard star and the GC object, respectively.}  
\end{minipage}
\end{figure*}

\begin{figure*}
\begin{minipage}{40pc}
\includegraphics[width=40pc,angle=0]{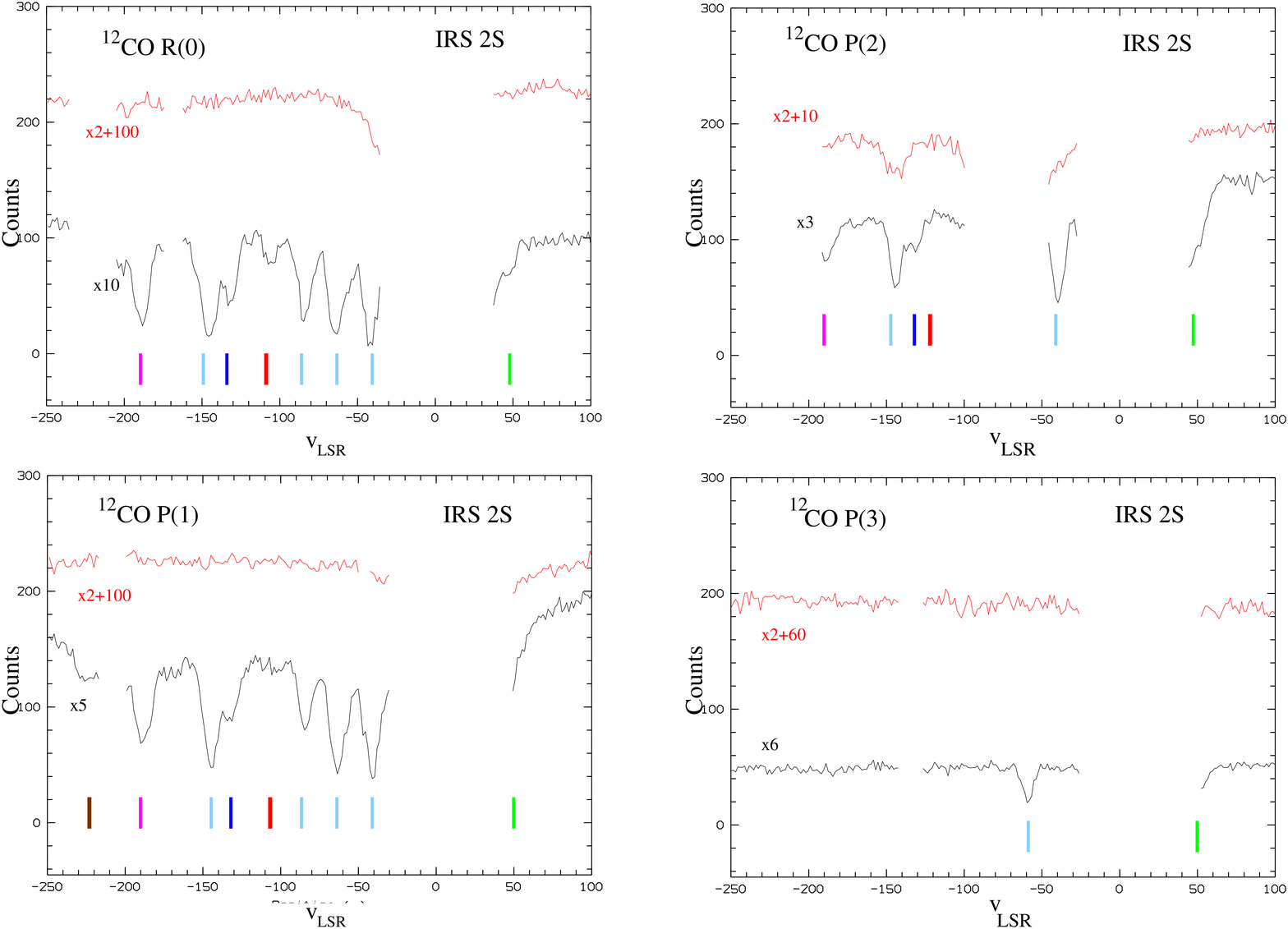} 
\caption{\label{IRS2S} Spectra (in black) of IRS 2S in the \CO12 R(0), P(1), P(2) and P(3) transitions. Also shown (in red) are  the spectra of the standard stars observed close to the science observation to correct for telluric lines. The blanked areas correspond to regions in the spectrum where the telluric corrections were not successful. The red arrows indicate the local absorptions at velocities listed in Table~\ref{tabvel}. The other colored arrows indicate the absorptions at the same velocities as in Figs.~\ref{sumR0_P1_P2} and \ref{sumR0P1P2}. Finally, the brown arrow indicates the line at -210~km/s. The scaling and/or shifting factors of the spectra are given in red and black for the standard star and the GC object, respectively.}  
\end{minipage}
\end{figure*}

\begin{figure*}
\begin{minipage}{40pc}
\includegraphics[width=40pc,angle=0]{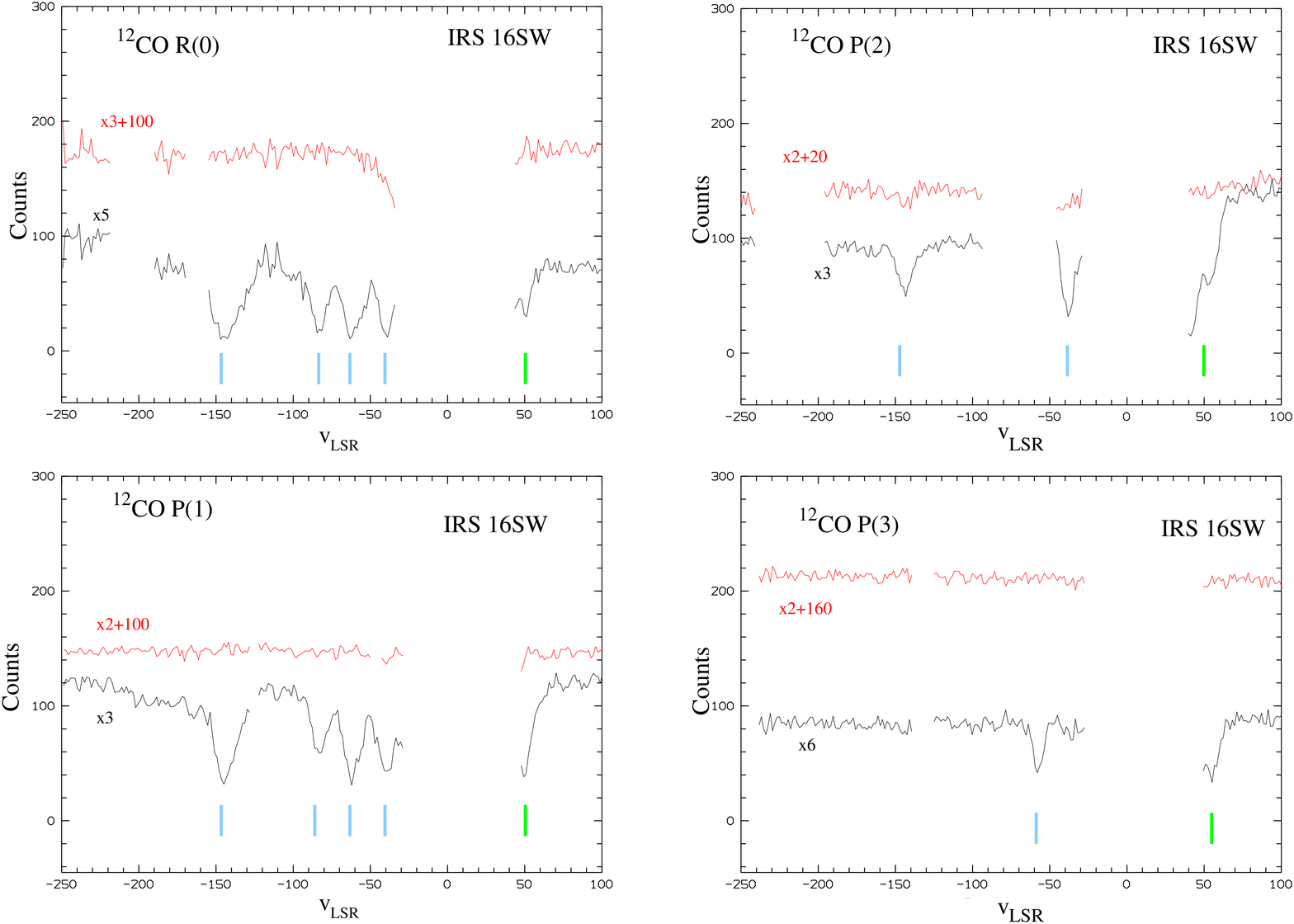} 
\caption{\label{IRS16SW} Spectra (in black) of IRS 16SW in the \CO12 R(0), P(1), P(2) and P(3) transitions. Also shown (in red) are  the spectra of the standard stars observed close to the science observation to correct for telluric lines. The blanked areas correspond to regions in the spectrum where the telluric corrections were not successful. The red arrows indicate the local absorptions at velocities listed in Table~\ref{tabvel}. The other colored arrows indicate the absorptions at the same velocities as in Figs.~\ref{sumR0_P1_P2} and \ref{sumR0P1P2}. The scaling and/or shifting factors of the spectra are given in red and black for the standard star and the GC object, respectively. } 
\end{minipage}
\end{figure*}

\begin{figure*}
\begin{minipage}{40pc}
\includegraphics[width=35pc,angle=0]{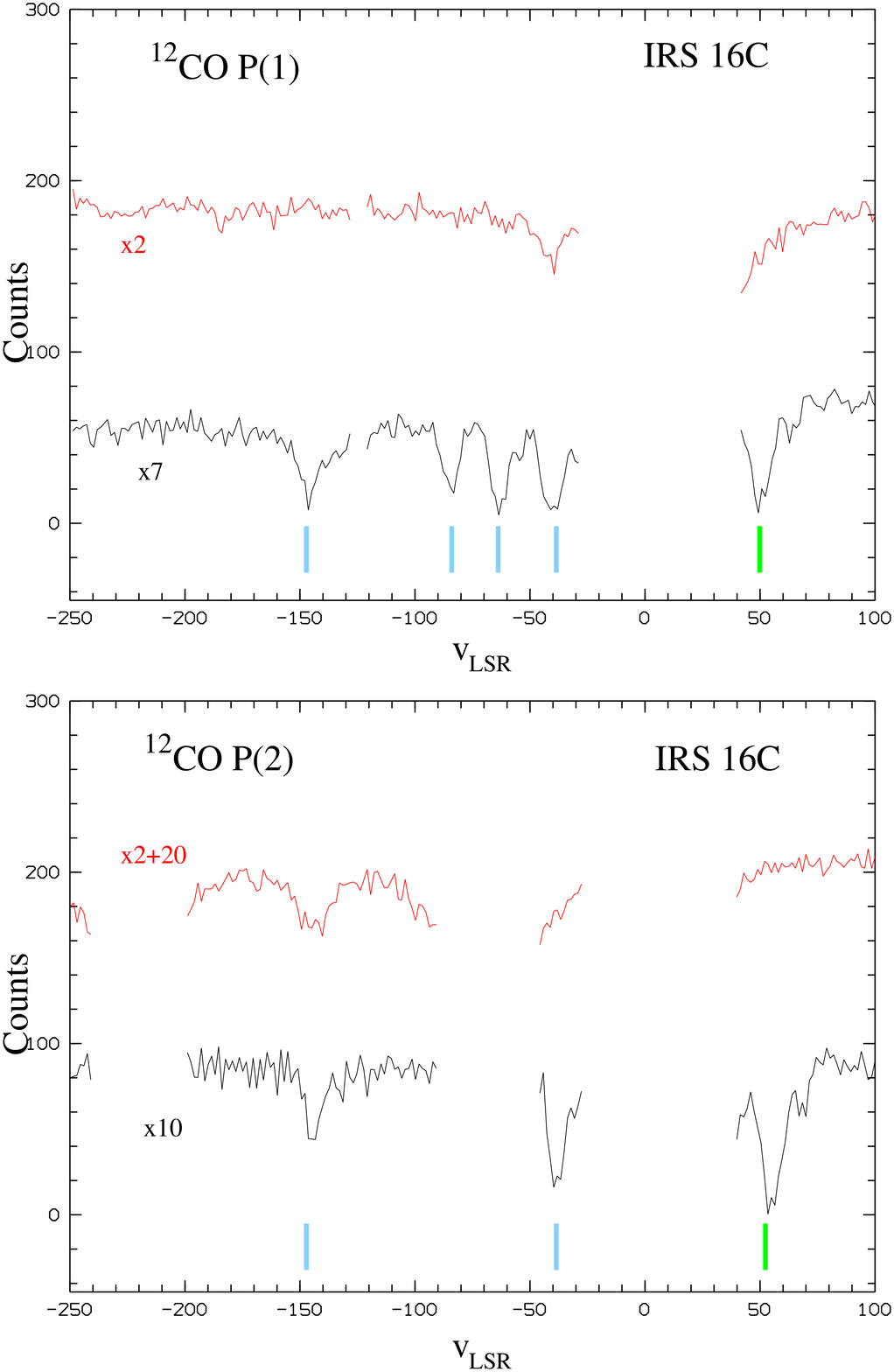} 
\caption{\label{IRS16C} Spectra (in black) of IRS 16C in the \CO12 P(1) and P(2) transitions. Also shown (in red) are the spectra of the standard stars observed close to the science observation to correct for telluric lines. The blanked areas correspond to regions in the spectrum where the telluric corrections were not successful. The red arrows indicate the local absorptions at velocities listed in Table~\ref{tabvel}. The other colored arrows indicate the absorptions at the same velocities as in Figs.~\ref{sumR0_P1_P2} and \ref{sumR0P1P2}. The scaling and/or shifting factors of the spectra are given in red and black for the standard star and the GC object, respectively.}  
\end{minipage}
\end{figure*}

\end{appendix}
\end{document}